\documentclass[lettersize,journal]{IEEEtran}
\usepackage[english]{babel}
\usepackage{blindtext}
\usepackage[caption=false,font=normalsize,labelfont=sf,textfont=sf]{subfig}
\usepackage[T1]{fontenc} 
\usepackage{amsmath,amsfonts}
\usepackage{amssymb}
\usepackage{algorithmic}
\usepackage{graphicx}
\usepackage{textcomp}
\usepackage{xcolor}
\usepackage{booktabs}
\usepackage{enumitem}
\usepackage[linesnumbered,ruled,vlined]{algorithm2e}
\usepackage{algorithmic}
\usepackage{xspace}
\usepackage{pdftexcmds}
\usepackage{etoolbox}
\usepackage{tikz}
\usepackage{ifsym}
\usepackage{tabularx}
\usepackage{amsthm}
\usepackage{multirow}
\usepackage{proof-at-the-end}
\usepackage{mwe}
\usepackage{makecell}
\usepackage{siunitx}

\usetikzlibrary{arrows,
                calc,
                positioning,
                quotes}
\usetikzlibrary{shapes.misc}
\usetikzlibrary{shapes.geometric}
\usetikzlibrary{patterns}
\usetikzlibrary{decorations.pathreplacing}
\usepackage[most]{tcolorbox} 

\tcbset {
  base/.style={
    arc=0mm, 
    bottomtitle=0.5mm,
    boxrule=0mm,
    colbacktitle=black!10!white, 
    coltitle=black, 
    fonttitle=\bfseries, 
    left=2.5mm,
    leftrule=1mm,
    right=3.5mm,
    title={#1},
    toptitle=0.75mm, 
  }
}

\definecolor{brandblue}{rgb}{0.34, 0.7, 1}
\newtcolorbox{mainbox}[1]{
  colframe=brandblue, 
  base={#1}
}

\newtcolorbox{subbox}[1]{
  colframe=black!30!white,
  base={#1}
}

\newcommand{\nodetikz}[1]{\begin{tikzpicture}[-,baseline=-0.6ex,main node/.style={circle,draw,font=\sffamily}]

\node[main node, fill={rgb:black,1;white,5},scale=0.55] (#1) {#1};\end{tikzpicture}}

\newcommand{\linktikz}[2]{\begin{tikzpicture}[-,auto,node distance=1cm,baseline=-0.6ex,main node/.style={circle,draw,font=\sffamily}]

\node[main node, fill={rgb:black,1;white,5},scale=0.55] (#1) {#1};
\node[main node, fill={rgb:black,1;white,5},scale=0.55] (#2) [right of = #1] {#2}; 
\path 
(#1) [-] edge node {} (#2);
\end{tikzpicture}}

\newcommand{\linktikzzz}[3]{\begin{tikzpicture}[-,auto,node distance=1cm,baseline=-0.6ex,main node/.style={circle,draw,font=\sffamily}]
  \node[main node, fill={rgb:black,1;white,5},scale=0.55] (#1) {#1};
  \node[main node, fill={rgb:black,1;white,5},scale=0.55] (#2) [right of = #1] {#2}; 
  \node[main node, fill={rgb:black,1;white,5},scale=0.55] (#3) [right of = #2] {#3}; 
  \draw[-] (#1) -- (#2);
  \draw[-] (#2) -- (#3);
  \end{tikzpicture}}

\newcommand{\linktikzzzz}[4]{\begin{tikzpicture}[-,auto,node distance=1cm,baseline=-0.6ex,main node/.style={circle,draw,font=\sffamily}]
  \node[main node, fill={rgb:black,1;white,5},scale=0.55] (#1) {#1};
  \node[main node, fill={rgb:black,1;white,5},scale=0.55] (#2) [right of = #1] {#2}; 
  \node[main node, fill={rgb:black,1;white,5},scale=0.55] (#3) [right of = #2] {#3}; 
  \node[main node, fill={rgb:black,1;white,5},scale=0.55] (#4) [right of = #3] {#4}; 
  \draw[-] (#1) -- (#2);
  \draw[-] (#2) -- (#3);
  \draw[-] (#3) -- (#4);
  \end{tikzpicture}}

  \newcommand{\linktikzzzzz}[5]{\begin{tikzpicture}[-,auto,node distance=1cm,baseline=-0.6ex,main node/.style={circle,draw,font=\sffamily}]
    \node[main node, fill={rgb:black,1;white,5},scale=0.55] (#1) {#1};
    \node[main node, fill={rgb:black,1;white,5},scale=0.55] (#2) [right of = #1] {#2}; 
    \node[main node, fill={rgb:black,1;white,5},scale=0.55] (#3) [right of = #2] {#3}; 
    \node[main node, fill={rgb:black,1;white,5},scale=0.55] (#4) [right of = #3] {#4}; 
    \node[main node, fill={rgb:black,1;white,5},scale=0.55] (#5) [right of = #4] {#5}; 
    \draw[-] (#1) -- (#2);
    \draw[-] (#2) -- (#3);
    \draw[-] (#3) -- (#4);
    \draw[-] (#4) -- (#5);
    \end{tikzpicture}}

\newtheorem{Definition}{Definition}

\newcommand{\ie}{\emph{i.e.,}\xspace}
\newcommand{\eg}{\emph{e.g.,}\xspace}

\def\preceqdot{\mathrel{\prec\kern-.5em\raise.05ex\hbox{$\cdot$}}} 
\def\lhddot{\mathrel{\lhd\kern-.35em\raise.035ex\hbox{$\cdot$}}} 
\def\lhdeq{\mathrel{\leqslant\kern-.498em\raise.255ex\hbox{$\shortmid$}}} 

\newcommand{\compRelationeqZ}{=_{\setminus 0}}

\newcommand{\N}{\mathbb{N}}

\newcommand{\relation}{\mathcal{R}}
\newcommand{\relationStrong}{\overset{\lozenge}{\mathcal{R}}}

\newcommand{\compRelation}{\preccurlyeq}
\newcommand{\compRelationZ}{\leqslant_{\setminus 0}}
\newcommand{\compRelationStrict}{\prec}
\newcommand{\compRelationStrictZ}{<_{\setminus 0}}

\newcommand{\extended}[1]{
    \overset{\lozenge}{#1}
}

\newcommand{\inPQ}[1]{
    \overset{#1}{\extendable}
}

\newcommand{\inPQRExtended}{%
\tikz[baseline]{
    \node[anchor=base] {\scriptsize $\relation$};
    \begin{scope}[yshift=0.2cm]
    \node[anchor=base] {\scriptsize $\lozenge$};
    \end{scope}
    \begin{scope}[yshift=-0.15cm]
    \node[anchor=base] {$\extendable$};
    \end{scope}
}%
}

\newcommand{\compRelationLex}{\lhdeq}
\newcommand{\compRelationStrictLex}{\lhd}

\newcommand{\compRelationAll}{%
\tikz[baseline]{
        \node[anchor=base] {$\compRelationStrict$};
        \begin{scope}[xshift=0.06cm]
        \node[anchor=base] {$\compRelationStrict$};
        \end{scope}
    }  
}
\newcommand{\compRelationAllLex}{%
\tikz[baseline]{
        \node[anchor=base] {$\lhd$};
        \begin{scope}[xshift=0.06cm]
        \node[anchor=base] {$\lhd$};
        \end{scope}
    }  
}

\newcommand{\extendable}{
    \hookrightarrow
}
\newcommand{\Encode}{\textnormal{\textsc{Encode}}}

\newcommand{\Seg}{\mathit{Seg}}
\newcommand{\problem}{\mathcal{P}}
\newcommand{\problemProperties}{\mathit{Properties}}

\newcommand{\Constraints}{\mathcal{C}}

\newcommand{\algo}{\ensuremath{\mathcal{A}}\xspace}

\newcommand{\looseEncoding}{\ensuremath{\textsc{Encode}}\xspace}
\newcommand{\Ssrc}[1][]{\ensuremath{S^{#1\mathit{src}}}\xspace}
\newcommand{\Sdest}[1][]{\ensuremath{S^{#1\mathit{dst}}}\xspace}
\newcommand{\Sdst}{\Sdest}
\newcommand{\Stype}[1][]{\ensuremath{S^{#1\mathit{type}}}\xspace}
\newcommand{\Adj}{\ensuremath{\mathit{Adj}}\xspace}
\newcommand{\Node}{\ensuremath{\mathit{Node}}\xspace}
\newcommand{\NodeAdj}{\ensuremath{\mathit{NodeAdj}}\xspace}
\newcommand{\NodeTypes}{\ensuremath{\mathit{SegTypes}}\xspace}

\newcommand{\LastSeg}{\ensuremath{\mathtt{LastSeg}}\xspace}
\newcommand{\NewLastSeg}{\ensuremath{\mathtt{NewLastSeg}}\xspace}

\renewcommand{\paragraph}[1]{\subsubsection*{#1}}

\begin{document}

\author{Quentin Bramas, Jean-Romain Luttringer, Pascal M\'erindol
\thanks{All authors are within the ICube laboratory, University of Strasbourg, France.}
}
\title{A Simple and General Operational Framework to Deploy Optimal Routes with Source Routing}


\maketitle

\begin{abstract}

  Source Routing, currently facilitated by Segment Routing (SR), enables precise control of forwarding paths by specifying detours (or \emph{segments}) to deviate IP packets along routes with advanced properties beyond typical shortest IGP paths. Computing the desired optimal segment lists, known as \emph{encoding}, leads to interesting challenges as the number of detours is tightly constrained for hardware performance. Existing solutions either lack generality, correctness, optimality, or practical computing efficiency -- in particular for sparse realistic networks. In this paper, we address all such challenges with GOFOR-SR. Our framework extends usual path computation algorithms to inherently look at optimal and feasible segment lists, streamlining the deployment of TE-compliant paths. By integrating encoding within the path computation itself and modifying the distance comparison method, GOFOR allows algorithms with various optimization objectives to efficiently compute optimal segment lists.
  Despite the loss of substructure optimality induced by SR, GOFOR proves particularly efficient, inducing only a linear overhead at worst. It also offers different strategies and path diversity options for intricate TE-aware load-balancing. We formally prove the correctness and optimality of GOFOR, implement our framework for various practical use-cases, and demonstrate its performance and benefits on both real and challenging topologies.

\end{abstract}


\section{Introduction}

\newcommand{\myparagraph}[1]{\paragraph{\textbf{\emph{#1}}}}

IP networks typically employ a best-effort, hop-by-hop routing paradigm. Packets follow paths minimizing the IGP cost, an additive metric considering link bandwidth, delay, or some operator design intent. While this approach offers some level of performance, it lacks more elaborated forwarding guarantees.

Despite the scalability advantages of considering uni-dimensional paths, some use cases require more robust or intricate routes. For instance, specific paths are necessary for circumventing failures (e.g., as computed by TI-LFA~\cite{ti-lfa}). Premium real-time flows may require specific paths ensuring guarantees both in bandwidth and latency (e.g., as computed by solving the Delay Constrained Least Cost problem, or DCLC~\cite{LUTTRINGER2022109015}).

Such paths may be deployed over best-effort ones through \emph{Source routing}. Source routing enables the source (\eg an edge provider router) to enforce paths with intricate properties. Packets are encapsulated to convey forwarding instructions to downstream routers, ensuring adherence to the desired path instead of best-effort routes. These instructions typically do not specify the path in its entirety, but encode it \emph{loosely} as a list of mandatory checkpoints that the packet must go through (following the shortest IGP paths between each checkpoint by default).

Despite its benefits, the (loose) source-routing paradigm encountered deployment challenges due to cumbersome troubleshooting and significant protocol complexity in existing control planes (e.g., RSVP-TE). Consequently, large-scale adoption of fine-grained Traffic Engineering (TE) through source routing was infrequent~\cite{srbook1}.

More recently, Segment Routing~\cite{rfc8402} (SR) introduced a scalable and lightweight implementation of the loose source-routing paradigm. SR sparked the interest from both academia and the industry. As of now, the majority of network operators deploy (or plan to deploy) SR for various use-cases, including Traffic-Engineering~\cite{SRusage}.

SR allows the source to prepend \emph{segments} to each packet. These segments typically represent nodes or links within the network, serving as mandatory \emph{detours} from the standard IGP paths. There are two main types of segments:

\begin{itemize}
    \item An adjacency segment designates a specific link that the packet should traverse.
    \item A node segment designates a particular node that the packet should pass through. Packets are forwarded to the node along the best path(s) on the IGP topology. SR supports multi-topologies~\cite{rfc8402}, allowing for example to follow least-delay sub-paths rather than IGP ones.
    \end{itemize}

\noindent

Various types of segments can be combined to encode any desired path. To deploy optimal paths with respect to the Traffic Engineering (TE) objective, it thus becomes necessary to compute the appropriate \textit{segment lists} and not only the paths themselves.


The computation of segment lists introduces an extra challenge: respecting the Maximum Segment Depth (MSD). This limit varies based on hardware capabilities, ranging from as few as 3 to up to 10 segments at line-rate~\cite{8251221}. 
Minimizing or at least controlling the number of segments when computing segment lists thus becomes critical.

In this paper we present our recipe to solve this problem in a generic and efficient manner. Given \emph{any} objectives, our framework computes the optimal segment lists respecting the MSD limit. Before presenting in detail our achievements, we first illustrate the challenges raised by the problem we tackle.

\myparagraph{Encoding optimal paths into segment lists is not enough}

Translating a given optimal path into a minimal segment lists, known as the \emph{path encoding problem}, leads to some algorithmic challenges. The encoding itself is arguably not the main one. However, encoding a (pre-computed) optimal path may lead to a segment list exceeding the MSD limit, as optimizing the number of segments may not align with optimizing the TE objective. This implies that path encoding must be performed \emph{during} the path computation,
in order to account for the number of segments when extending paths.

This phenomenon is familiar within the context of computing multi-metric constrained paths. Considering the number of segments as an additional metric, it is well-known that, to find solutions satisfying a constraint on a metric, all \emph{non-dominated} paths must be extended for optimality correctness\footnote{In a multi-criteria context, a  path is \emph{optimal} (or \emph{non-dominated}) if it is better on at least one metric than any other path towards the same node. Intuitively, a dominated path, being worse on all metrics, cannot become optimal later on and may be pruned from the exploration.}.
However, it works only if the metrics are isotone (\ie an optimal path is composed of optimal subpaths), but \textbf{this fundamental property does not hold when considering source routing}. This issue being at the core of our contribution, we precisely illustrate it in the following paragraph.

\myparagraph{Illustration of the challenges: MSD constraint \& Isotonicity} \label{sec:illus}

We now show why ignoring some dominated distances may lead to incorrect (possibly infeasible, \ie non-deployable) solutions.

We illustrate this challenge motivating the use of GOFOR-SR on a basic use-case (Fig.~\ref{fig:test}). The objective is to minimize the delay (called \emph{Least-Delay} in this paper). We restrict our analysis to adjacency and IGP node segments.
Fig.~\ref{fig:test} shows a multi-valuated graph, where each link exhibits both an IGP cost and a delay. Possible distances to reach nodes \nodetikz{3} and \nodetikz{D} are shown below said nodes, exhibiting the IGP cost, the delay (and the required number of segments). The associated segment lists can be seen on the right side of their respective distances. 

The first observation is that reaching \nodetikz{D} with a delay at most 4 requires at least 3 segments, but if MSD is set to 2 at \nodetikz{S}, the orange path has to be chosen. Pre-computing the blue or the green optimal paths (delay-wise) and applying an encoding a posteriori would result in segment lists exceeding a tight MSD constraint. The orange segment list, while being dominated by the blue segment list at node \nodetikz{3}, becomes the best option with at most 2 segments to reach \nodetikz{D}. 

\tikzset{cross/.style={cross out, draw=red, minimum size=2*(#1-\pgflinewidth), inner sep=0pt, outer sep=0pt, fill=white, line width=0.7mm},
cross/.default={11pt}}

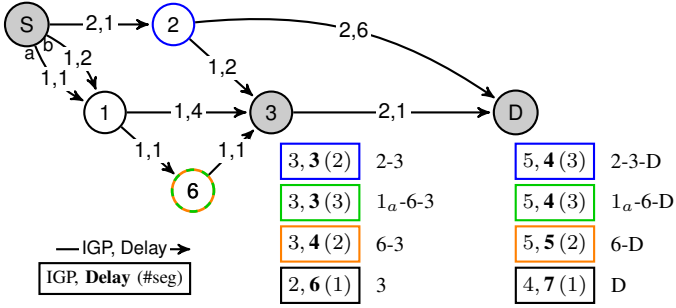
\begin{figure}
    \centering
           
\begin{tikzpicture}[yscale=1.3, xscale=1.3, -,>=stealth',shorten >=1pt,auto,node distance=3cm,
    thick,main node/.style={circle,fill=white,draw,font=\sffamily,inner sep=3pt},
    edgelabel/.style={fill=white,inner sep=1pt, anchor=center, pos=0.5,font=\sffamily}]

\begin{scope}[xshift=-1cm, yshift=0.1cm]\footnotesize
\node[main node, fill=black!20] (S) at (-0.5,1.7) {S};
\node[main node] (1)                at (0.3,0.8) {1};
\node[main node,line width=1pt, draw=orange] (6)                at (1.2,0) {6};
\node[circle,line width=1pt,draw,font=\sffamily,inner sep=3pt, draw=black!20!green,dashed] (6)                at (1.2,0) {6};
\node[main node,draw=blue] (2)                at (1,1.7) {2};
\node[main node,fill=black!20] (3)  at (2,0.8) {3};
\node[main node,fill=black!20] (D)  at (4.5,0.8) {D};
\node[font=\sffamily] at (-0.47,1.4) {\scriptsize a};
\node[font=\sffamily] at (-0.28,1.5) {\scriptsize b};

\draw (S) [->, bend left=20] edge[] node[edgelabel] {1,2} (1);
\draw (S) [->, bend left=-20] edge[] node[edgelabel] {1,1} (1);

\draw (1) [->] edge[] node[edgelabel] {1,1} (6);
\draw (1) [->] edge[] node[edgelabel] {1,4} (3);
\draw (6) [->] edge[] node[edgelabel] {1,1} (3);

\draw (3) [->] edge[] node[edgelabel] {2,1} (D);

\draw (S) [->] edge[] node[edgelabel] {2,1} (2);
\draw (2) [->] edge[] node[edgelabel] {1,2} (3);
\draw (2) [->, bend left=20] edge[] node[edgelabel] {2,6} (D);

\end{scope}

\footnotesize
\begin{scope}[xshift=-0.8cm]
\node[rectangle, draw=blue] (d1) at (2.3,0.4) {$3,\textbf{3}\,(2)$};
\node[rectangle, draw=black!20!green] (d2) [below=0.55 of d1.west,anchor=west] {$3,\textbf{3}\,(3)$};
\node[rectangle, draw=orange] (d3) [below=0.55 of d2.west,anchor=west] {$3,\textbf{4}\, (2)$};
\node[rectangle, draw] (d4) [below=0.55 of d3.west,anchor=west] {$2,\textbf{6}\, (1)$};
\end{scope}
\node[] (d1-d) [right=0.1 of d1]  {2-3};
\node[] (d2-d) [below=0.55 of d1-d.west,anchor=west] 
{$1_a$-6-3};
\node[] (d3-d) [below=0.55 of d2-d.west,anchor=west] 
{6-3};
\node[] (d4-d) [below=0.55 of d3-d.west,anchor=west] 
{3};

\begin{scope}[xshift=0.7cm]
\node[rectangle, draw=blue] (d11) at (3.2,0.4) {$5,\textbf{4}\, (3)$};
\node[rectangle, draw=black!20!green] (d12) [below=0.55 of d11.west,anchor=west] 
{$5,\textbf{4}\, (3)$};
\node[rectangle, draw=orange] (d13) [below=0.55 of d12.west,anchor=west] 
{$5,\textbf{5}\, (2)$};
\node[rectangle, draw] (d14) [below=0.55 of d13.west,anchor=west] 
{$4,\textbf{7}\, (1)$};
\end{scope}

\node[] (d21) [right=0.1 of d11]  {2-3-D};
\node[] (d22) [below=0.55 of d21.west,anchor=west] 
{$1_a$-6-D};
\node[] (d23) [below=0.55 of d22.west,anchor=west] 
{6-D};
\node[] (d24) [below=0.55 of d23.west,anchor=west] 
{D};

\draw (-1.2,-0.5) edge[->] node[edgelabel] {\scriptsize IGP, Delay} (0.2,-0.5);
\node[rectangle, draw] at (-0.6,-0.8) {\scriptsize IGP, \textbf{Delay} (\#seg)};

\end{tikzpicture}
    \caption{A Least-Delay use case highlighting  the limit of an \emph{a posteriori} encoding scheme and the loss of isotonicity. The orange path may be pruned at node 3 despite it becoming optimal if MSD is set to 2 on S.}
\label{fig:test}
\end{figure}

In general, distances being suboptimal at an intermediate node are not supposed to be extended further (as the latter should not possibly lead to optimal distances towards downstream nodes). This fundamental property, called \emph{isotonicity} or subpath optimality, is \emph{essential} to bound the worst-case complexity of the path computation (as only a manageable number of distances are extended).

In Fig.~\ref{fig:test}, when exploring the graph, this standard behavior would thus dictate to only extend the blue distance from node \nodetikz{3} onward. Indeed, the orange distance is dominated as it has a higher delay while not offering better options than the blue distance on other metrics. The green distance has an optimal delay, but also require more segments than its blue counterpart.

The surprising effect occurs when extending these intermediate distances by the edge \linktikz{3}{D}: the blue segment list now requires an additional segment to encode the desired distance while it is not the case for the two others. Starting from the current detour \nodetikz{2}, adding a single node segment to \nodetikz{D} does not encode the distance exhibited by the path \linktikzzzz{S}{2}{3}{D}, as flows will follow the link \linktikz{2}{D}, which exhibits a higher delay. Thus, while encoding the distances of the path \linktikzzz{S}{2}{3} required 2 segments, encoding the distances of the path \linktikzzzz{S}{2}{3}{D} requires one more segment.

On their sides, the green and orange segment lists do not evolve in the same fashion: they do not require more segments to encode the additional edge. Let us consider the green path: while encoding the distances of the path \linktikzzzz{S}{1}{6}{3} already required 3 segments, encoding the extended path \linktikzzzzz{S}{1}{6}{3}{D} just requires to modify the last node segment (\nodetikz{3}) by a node segment targeting \nodetikz{D}. 

The fact that some (but not all) segment lists require an additional segment when extended implies that at least some dominated segment lists \emph{should} be extended, as it is the case for the orange one in the example. Moreover, if one aims at retrieving \emph{all} optimal segment lists (\eg to perform TE-aware load balancing), the green segment list should also be extended, as it becomes equivalent to the blue one after the extension by the edge \linktikz{3}{D}.

This observation bears a drastic consequence: it seems that to properly \emph{wrap} SR around existing algorithms (\ie augmenting them to compute minimal and optimal segment lists), \emph{all} paths should be extended to ensure that the optimal solution is found. This challenge leads to the main research question of this paper: 

\begin{subbox}{Research Question}
    \textbf{How can we extend path computation methods} (with additive metrics and isotone properties), \textbf{to correctly, optimally and \emph{efficiently} wrap SR around them}?
\end{subbox}

Compared to the related works, we manage to answer this research question while remaining efficient and avoiding the use of heavy data-structures and graph transformation.

\myparagraph{Main Achievements}

Despite the practical interest of this research question, there exists a clear gap in the literature for generic means to augment path computation algorithms with the ability to consider SR both properly and efficiently.

Our answer is \textbf{GOFOR}-SR, a \textit{\textbf{G}eneral \textbf{O}perational \textbf{F}ramework for \textbf{O}ptimal \textbf{R}outes} with SR. GOFOR is a simple and generic framework that enables path computation algorithms to efficiently compute optimal segment lists. Despite the complexity of the problem, GOFOR allows their retrieval with linear overhead at worst. As the first generic and optimal solution for generic segment lists computation, GOFOR not only paves the way for easy deployments of SR in various use cases, but also for intricate TE-aware load balancing.

Our contributions are as follows:

\begin{enumerate}
    \item The formalization of a new SR model for the problem(s) we tackle. We introduce the notions of path encoding, \textbf{SR wrapping (constrained or lexicographical), multi-topology segment lists, and path diversity strategies} in Sec.~\ref{sec:notations};
    
    \item A loose, ECMP-friendly, encoding scheme able to compute minimal segment lists possessing a given set of properties. This encoding is the first to handle multi-topologies, and is performed with a linear time overhead (wrt. \#nodes) during path exploration (Sec.~\ref{sec:encoding}); 

    \item \textbf{An extended dominance function in order to re-ensure an isotone relation} (Sec.~\ref{sec:iso}). We define a new, SR-aware dominance to correctly compare segment lists. This relation can be implemented within any traditional shortest path algorithm along with our encoding scheme to ensure that the optimal deployable segment lists are found for only a linear overhead (wrt. \#nodes) (Sec.~\ref{sec:recipe}).

    \item A C implementation of GOFOR and three modules showing how our framework can be used to properly wrap SR around algorithms computing $(i)$ Delay Constrained Least Cost paths, $(ii)$ Least Delay paths, and $(iii)$ Best IGP paths avoiding a link failure. We experimentally show that GOFOR-SR allows reaching better performance than concurrent approaches (Sec.~\ref{sec:usecase}). The code of our implementation is available online~\cite{sourcecode} and a python notebook is provided to ease the reproducibility of our experiments~\cite{sourcecode}.
\end{enumerate}

\section{The SR-Wrapped Problem in a Formal Nutshell}
\label{sec:notations}

\subsection{Notations \& Models}

Let $G(E, V, w)$ be a directed graph with $V$ the set of vertices, $E$ the set of edges, and $w : \N_+^k$ as the weight function mapping each edge to its weight vector $(w_1(e), \ldots, w_k(e))$. Observe that initial metrics are indexed from 1 to $k$; metric with index $1$ denotes the IGP distance -- while, as formally defined later, metric with index $0$ will represent the number of segments. Other components may be any desired additive metrics. For a path $p$, $p[i]$ denotes the $i$-th traversed node of $p$ and $p[i: j]$ denotes the subpath of $p$ from $p[i]$ to $p[j]$.

\myparagraph{Segments}
A segment $S=\Seg(\Stype, \Ssrc, \Sdest)$ represents a set of paths from a source \Ssrc to a destination \Sdest. The set depends also on its type \Stype, defining the kind of segment in use (either directly an adjacence or a subpath satisfying a distance regarding a given metric). If no path exists from $\Ssrc$ to $\Sdst$ with the property associated with $\Stype$, we say that the segment is empty. 
Let $S$ be a segment and $e = (\Sdst, v)$ an edge. We say that $S$ can be extended by $e$ if the segment $\Seg(\Stype, \Ssrc, v)$ is not empty and contains a path (of type \Stype) from $\Ssrc$ to $v$ passing through $e$. We denote by $S \circ e = \Seg(\Stype, \Ssrc, v)$ this extension of $S$ by $e$, otherwise we write $S \circ e = \emptyset$. 

A segment satisfies two properties: (a) if $p$ is a path in $S$, then its sub-segments are non-empty: $\forall i<j, (\Stype, p[i], p[j])\neq\emptyset$; and (b) if a path $p$ is in a segment $S$ and $S\circ e$ is not empty, then $p\circ e$ is in $S\circ e$.

The most common types of segments are \emph{adjacency}, denoted $\Adj$, and \emph{node}, denoted $\Node$.
If $\Stype = \Adj$, $S$ simply represents the edge $(\Ssrc, \Sdest)$ if it exists, and is empty otherwise. By construction, an adjacency segment cannot be extended, so it verifies the two properties above.

If $\Stype = \Node_i$, $S$ represents the set of paths from $\Ssrc$ to $\Sdest$ minimizing the $i$-th metric $w_i$. In this context, $\Node$ stands for ``all the best paths towards the destination \Sdest''. 
Node segments satisfy the two properties above because sub-paths of shortest paths are also shortest paths.

In this paper, and in particular in our examples, we mainly consider segments with types $\Adj$ and $\Node_1$ (IGP based node segments as they are the only ones provided by default with SR), but our recipe works for any type of segments and their combination. 
GOFOR can also support IPv6 segments through the $\NodeAdj_i$ segment type~\footnote{In SR-MPLS, guiding a packet through a link $(u,v)$ involves two segments: first, a node segment to direct the packet to $u$, then followed by the relevant adjacency segment (whose local significance varies depending on the interpreting node). With SRv6, accomplishing this task requires just a single adjacency segment as the local ports of each node are globally broadcasted.}. A $\NodeAdj_i$ segment between $u$ and $v$ refers either $(i)$ to the $(Node_i, u, v)$ segment if it is not empty, or $(ii)$ to all the paths obtained by concatenated a path in $(Node_i, u, v')$ with the edge $(v',v)$, for an arbitrary neighbor $v'$ of $v$. If no such neighbor exists, then the segment is empty. 
In the first case, the segment satisfies its two required properties because it is a node segment. In the second case, the segment is not extendable\footnote{Otherwise it implies that the corresponding node segment is non-empty and thus contradicts the fact that the second case is considered.}, so it also satisfies the two properties (a and b). 
The set of types of segments (\ie $\Node_i$, $\Adj$ and $\NodeAdj_i$) is denoted $\NodeTypes$.

\myparagraph{Distances}
We assume all the metrics are additive. Therefore, the distance $d(p)$ of a path $p$ is simply the sum of the weights of its edges, \ie for each metric $i$ we have $d_i(p)=\sum_{e\in p} w_i(e)$.

As a segment $S$ represents a set of paths, its distance $d(S)$ is defined as the maximum distance among all paths in $S$ \emph{for each metric}. In other words, $d_i(S) = \max_{p\in S} d_i(p)$. Note that $d(S)$ 
may not correspond to the distance of a specific path in $S$, since the maximum value for each metric might not be attained by the same path. Furthermore, observe that all the paths in a segment $\Seg(Node_i, u, v)$ have the same distance for the metric $i$, since, by definition, they all minimize this metric between $u$ to $v$. Finally, we denote $\Gamma$ the size of the Pareto Front (\ie the number of non-dominated paths) induced by the set of the $k$ metrics (at least with indexes 0 and 1).

\myparagraph{Segment Lists}

A segment list $L$ is defined as a finite sequence of non-empty segments $L=(S_1, S_2, \ldots, S_l)$. In a segment list, each segment $S_i$ starts at the end of the previous segment: more formally, for all $1 < i \leq l$, $\Sdest_{i-1} = \Ssrc_i$. 

The distance $d(L)$ of a segment list $L$ is the sum of the distances of its segments, \ie for each metric: $d_i(L) = \sum_{j=1}^l d_i(S_j)$.
Moreover, we consider the number of segments as the 0-th metric, so that $d_0(L) = l$ ($d_0$ not being defined for paths). Note that $d_0(L)$ must be lower than $MSD$ if the number of segments is constrained. A segment list distance $d(.)$ is a vector in $\N_+^{k+1}$.

We say that a segment list $L$ can be edge-extended by an edge $e$ if $e$ can be concatenated to the last segment of $L$ \ie $S_l\circ e$ is not empty. In this case the resulting segment list is denoted $L\circ e = (S_1, S_2, \ldots, S_l\circ e)$.
Two segment lists $L_1$ and $L_2$ can be concatenated if the destination of the last segment in $L_1$ matches the source of the first segment in $L_2$. This concatenation is denoted $L_1\oplus L_2$.

\myparagraph{Two Encoding Paradigms}

Encoding a path $p$ consists in finding a list of detours encoding paths exhibiting properties of which $p$ is a representative. We define two types of encoding.

\begin{Definition}[Strict Encoding]
We say a segment list $L$ strictly encodes a path $p$ if $L$ is a partition of $p$. In other words, each segment in $L$ contains only one subpath of $p$ and $L$ has the same source and destination as $p$. 
\end{Definition}

\begin{Definition}[Loose Encoding]
We say that a segment list $L$ loosely encodes a path $p$ if each segment in $L$ contains a subpath of $p$, $L$ has the same source and destination as $p$, and $d(L) = d(p)$ (in this equality we ignore the number of segments, since it is not defined for paths).
\end{Definition}

Strict encoding is useful when considering use-cases such as monitoring or services chaining, in which the structure of the path must be enforced. While this type of encoding is commonly found in the literature \cite{SCMon, 7778603}, loose encoding is better suited for most usual use-cases as it enables shorter segment lists benefiting from multiple load balanced paths (ECMP) having all bounded guarantees.

On the contrary to the strict paradigm, a packet that is source-routed through the loose encoding of a path $p$ may not follow $p$ (or just partially) but its effective route will have the same (or better) distance(s). For example, when relying on the loose paradigm to encode a least-delay path $p$ (typically without using node segments of this type), this encoding ensures that any path in the resulting segment list $L$ will possess the same delay as $p$, or a better one. Note that $p$ remains within the set of paths encoded by $L$.

This paradigm is the most suited when the distance(s) matter more than the structure of the path itself, which is often the case in practice, and enables to mitigate $d_0(L)$. Note that topological properties (\eg avoiding a failed component) can also be enforced, and many other usecases can be enforced too (\eg applying a given instruction on a given node such as a firewall -- or just going through an arbitrary node).

\myparagraph{The SR Graph}

SR-aware routers need network-wide knowledge of available segments to build paths. As segments encode shortest paths, computing all possible segments requires running an All-Pair-Shortest-Path algorithm.

The resulting segment database can be organized as a graph, commonly referred to as the SR Graph, where an edge $(u,v)$ represents either a node segment $(Node_1, u, v)$ or an adjacency segment $(Adj, u, v)$ (the edge weights being set as $d(S)$ within the graph)~\cite{LUTTRINGER2022109015,AubryPhD}.

Associated algorithms usually explore the SR Graph directly, exploiting the fact that paths computed on this graph are, in fact and by design, segment lists. This eliminates the need for a loose conversion algorithm to convert paths to segment lists, and allows them to terminate exploration once paths exceed $MSD$ hops.

However, inherently if originally connected, the SR Graph is fully-meshed and so has a quadratic number of edges ($|E|=|V|^2$), significantly denser than the underlying initial network graph (where $|E|$ is generally similar to $|V|$ up to a multiplicative scalar factor). Exploring the SR Graph is costly and requires specialized algorithms not easily generalized to all use-cases.

GOFOR takes a different approach. It doesn't mandate the algorithms it extends to handle the complete density of an SR Graph. Instead, SR-wrapped algorithms can still explore the original (sparse) network graph. GOFOR uses the SR Graph indirectly as an efficient lookup table to facilitate the conversion of paths to segment lists. This approach enables our framework to be the most efficient in realistic cases as it leverages the sparsity of the initial network graph.

\subsection{Problem Statement: How to SR-Wrap a given Path Computation Algorithm ?}

Our main goal with GOFOR is to adapt an algorithm capable of solving a given (non-SR) problem to its SR-related version. We call this process \emph{SR-wrapping} as it adds the SR metric (\ie the number of segments) and MSD constraint to the initial problem.
Generally, a path computation problem is defined by a set of (optional) properties (\eg avoiding a link or a node), (optional) constraints that the solutions should verify (\eg keeping the delay under a given threshold), plus an objective function that the operator aims to optimize (\eg minimizing the delay).

\myparagraph{Properties, Initial Constraint(s) and Objective(s)}

We define the set of properties as a predicate $\problemProperties$. Considering a graph $G$ and a destination $v$, $\problemProperties$ takes as input a path $p$ and returns \emph{true} if $p$ has destination $v$ and is a valid solution to the problem in $G$. For a segment $S$, $\problemProperties(S)$ is true if the predicate is verified $\forall p \in S$. We assume $\problemProperties$ to be isotonic: if $\problemProperties(p)$ is true, then $\problemProperties(p')$ is true for any subpath $p'$ of $p$. The typical example we have in mind is: the sequence of edges in $p$ does not include any failed components.

The initial constraints, if any, are defined as a vector $\Constraints_{\setminus 0} = (c_1,\ldots, c_k)$, where each $c_i$ sets an upper bound on the distance $d_i$ that must be satisfied for all returned path $p$. Formally, we have $d(p) < \Constraints_{\setminus 0}$. Only minimizing a given $d_i$, without constraining it, is equivalent to set $c_i = \infty$.

The objective function can be defined as a \textit{comparison relation} $\compRelationZ$, that applies to the distances, hence define the paths' distances ranking in the initial problem. The relation $\compRelationZ$ could either represent a totally ordered relation with a given lexicographical ranking (\eg sorting paths according to their delay or cost, or both one after the other), or more intricate partially ordered relations such as Pareto-optimality (inducing a non-trivial Pareto front, \ie $\Gamma > 1$), in particular when considering multi-constrained problems.

Overall, $\compRelationZ$ denotes the initial ordered relation (partial or total)\footnote{More precisely, any order relation that contains the component-wise order \ie $\forall i, x_i\leq y_i\Rightarrow x\compRelationZ y$. We assume monotone and isotone relations by construction (additive strictly increasing metrics because of the nature of $w$).}
, while $\compRelationStrictZ$ denotes its associated strict counterpart. 

\myparagraph{Towards the SR-Wrapped Problem}

From the inputs described above -- covering most of the basic path computation problems -- we will now define $\problem$, as a problem consisting in finding the \emph{optimal segment lists} compliant with the initial problem characteristics (properties, constraints and objectives). 

Several steps are necessary to formulate such an SR-wrapped problem.
While the set of $\problemProperties$ does not need to be modified (as considering SR does not affect them), the constraints $\Constraints_{\setminus 0}$ have to be extended to $\Constraints = (c_0, c_1,\ldots, c_k)$, where $c_0 = MSD$ represents an (optional) upper bound on the number of segments. 
More importantly, the comparison relation, $\compRelationZ$, also has to be modified to consider the number of segment as a new metric: the goal is now to compare the distances of segment lists (rather than paths').

\myparagraph{Two Main Strategies to Extend $\compRelationZ$}

There exist several ways to wrap SR around $\problem$, leading to different SR-wrapped variants of $\compRelationZ$. The two most relevant strategies are to consider the number of segments either in a \emph{lexicographical} or in a \emph{constrained} fashion. With a \emph{lexicographical} wrapped comparison, one aims at finding the minimal segment list(s) among the distance(s) optimizing the initial objective(s) (defined by $\compRelationZ$, under constraints $\Constraints_{\setminus 0}$ and verifying $\problemProperties$). MSD is thus technically ignored, and $d_0$ is minimized as a secondary objective (meaning that the results may exceed MSD).

The \emph{constrained} strategy aims to return a \emph{deployable} segment list(s) $L$ (\ie verifying $d_0(L) < c_0 = MSD$) whose underlying paths are optimal with respect to the initial problem. This method ensures that the returned segments lists are deployable, although the TE objective may have to be relaxed to find such a solution as 
a segment list that is not optimal with respect to the initial problem may become the only feasible solution for a subsequent destination.

Table~\ref{tab:allrel} shows the formal definitions of the relations associated to these strategies. Several relations can be derived from a single strategy, depending on the chosen \emph{path diversity option}. 

\begin{figure}[!ht]
    \begin{tabularx}{\linewidth}{l|l|l}
  Option&$\;\relation$ & Segment list $x$ $\relation$-dominates $y$
  , iff: \\
     \hline
     \multicolumn{3}{l}{\emph{Constrained-SR}}\\
     \hline
     oneBest&$\;\compRelation$ & $d(x) \compRelationZ d(y) \land  d_0(x) \leq d_0(y)$ \\
     allBest&$\;\compRelationStrict$ & $x \compRelation y \land d(x) \neq d(y)$ \\
     all&$\compRelationAll$ & $d(x) \compRelationStrictZ d(y) \land d_0(x) \leq d_0(y)$\\
     \hline     
     \multicolumn{3}{l}{\emph{Lexicographic-SR}}\\
       \hline  
       oneBest&$\;\compRelationLex$ & \makecell[l]{$d(x) \compRelationStrictZ d(y) \lor$\\$\qquad\left( d(x) \compRelationeqZ d(y) \land d_0(x) \leq d_0(y)\right)$}\\
       allBest&$\;\compRelationStrictLex$ & \makecell[l]{$d(x) \compRelationStrictZ d(y) \lor $\\ $\qquad\left( d(x) \compRelationeqZ d(y) \land d_0(x) < d_0(y)\right)$}\\
       all&$\compRelationAllLex$ & $d(x) \compRelationStrictZ d(y) $ \\
       \hline 
    \end{tabularx}
  \caption{Definitions of the set of relations supported by GOFOR. Each corresponds to a given SR-wrapping strategy and path diversity option.}
  \label{tab:allrel}
  \end{figure}

\myparagraph{And Three Options for Path Diversity}

GOFOR supports various path diversity options, in order to adapt to the operator's needs, \eg regarding load-balancing. 
Each of these options are applicable to both the lexicographical and the constrained strategies. 
We call the three diversity options \emph{1best}, \emph{allBest} and \emph{all}. With respect to the initial problem and the chosen strategy; \emph{1best} returns at least one optimal solution, \emph{allBest} returns all optimal solutions, and \emph{all} returns all solutions that are encodable with less than $c_0$ segments. The two last options allow the operator to perform source-routed load-balancing. 

  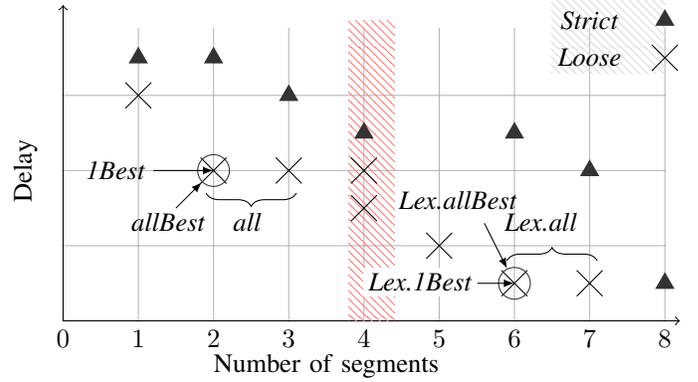
\begin{figure}
    \tikzset{cross/.style={cross out, draw=black, minimum size=3*(#1-\pgflinewidth), inner sep=0pt, outer sep=1pt},
    cross/.default={1pt},
    triangle/.style = {fill=black!80, regular polygon, regular polygon sides=3, inner sep=1.5pt}}
    \begin{tikzpicture}[scale=1]
      \draw[very thin,color=black!30] (-0.01,-0.01) grid (8.01,3.9);
      \draw[->] (-0,0) -- (8.2,0);
      \draw[->] (0,-0) -- (0,4.2);
      \node[rotate=90] at (-0.5,2) { Delay};
      \node at (3.5,-0.6) { Number of segments};
      \foreach \x in {0,...,8}
        \draw (\x,0) node[below] { $\x$};
    
      \fill[pattern=north west lines, pattern color=red!50] (3.8,0) rectangle (4.4,4);
    
      \draw (1,3) node[cross=3pt] {};
      \draw (2,2) node[cross=3pt] {};
      \draw (3,2) node[cross=3pt] {};
    
      \draw (4,2) node[cross=3pt] {};
    
      \draw (4,1.5) node[cross=3pt] {};
      \draw (5,1) node[cross=3pt] {};
      \draw (6,0.5) node[cross=3pt] {};
      \draw (7,0.5) node[cross=3pt] {};

      \draw (1,3.5) node[triangle] {};
      \draw (2,3.5) node[triangle] {};
      \draw (3,3) node[triangle] {};
      \draw (4,2.5) node[triangle] {};
      \draw (6,2.5) node[triangle] {};
      \draw (7,2) node[triangle] {};
      \draw (8,0.5) node[triangle] {};
    
      \fill[pattern=north west lines, pattern color=gray!27] (6.5,3.3) rectangle (8,4.3);
      \node at (7,4) { \emph{Strict}};
  
      \draw (8,4) node[triangle] {};
  
      \node at (7,3.5) { \emph{Loose}};
  
      \draw (8,3.5) node[cross=3pt] {};
    
      \draw [decorate,decoration={brace, mirror,amplitude=5pt}]
      (1.9,1.7) -- (3.1,1.7);
      \node at (2.45,1.32) { \emph{all}};
    
      \node[circle, draw=black!80, inner sep=0pt, minimum size=12pt] (allbest) at (2,2) {};
      \draw[-latex] (1.5,1.5) -- (allbest);
      \node at (1.4,1.32) { \emph{allBest}};
    
      \draw[-latex] (1.2,2) -- (2,2);
      \node at (0.7,2) { \emph{1Best}};
    
    \begin{scope}[shift={(4,-1.5)}]
      \draw [decorate,decoration={brace, amplitude=5pt}]
      (1.9,2.3) -- (3.1,2.3);
      \node at (2.35,2.82) { \emph{Lex.all}};
    
      \node[circle, draw=black!80, inner sep=0pt, minimum size=12pt] (allbest) at (2,2) {};
      \draw[-latex] (1.6,2.8) -- (allbest);
      \node at (1.25,3.1) { \emph{Lex.allBest}};
    
      \draw[-latex] (1.4,2) -- (2,2);
      \node[fill=white,inner sep=2pt] at (0.75,2.02) { \emph{Lex.1Best}};
    \end{scope}
    
    \end{tikzpicture}
    \caption{Optimal distances towards a given destination, depending on the SR-wrapping strategy (constrained or lexicographic), the path diversity option (1best, allBest or all) and the encoding scheme (loose or strict).}\label{fig:plop}
  \end{figure}

All the resulting relations are shown in Table~\ref{tab:allrel}. 
The desired form of the relation, noted $\relation$, is to be chosen according to the operator's wishes and based on the initial relation $\compRelationZ$. Solutions that are not optimal with respect to $\relation$ are said to be $\relation$-dominated. 

Figure~\ref{fig:plop} illustrates graphically the solutions returned by each strategy and option. We consider the minimization of the delay as the initial objective (using IGP node segments). For a given (hypothetical) destination, each cross (resp. triangle) represents a non-$\relation$-dominated, loose-encoded (resp. strict-encoded), segment lists. The y-axis shows the delay of the computed segment lists while the x-axis shows the required number of segments considering $MSD = 4$ (excluded). The \emph{constrained} options return only segment lists that are below the MSD constraint. 
  On the contrary, the \emph{lex} strategies only focus on the best delays.
  With the \emph{all} case, GOFOR returns either all optimal distances encodable in fewer than $4$ segments (with the \emph{constrained} strategy), or all optimal distances with respect to $\compRelationZ$ when considering the \emph{lex} strategy. 
  
Each pair (strategy, option) corresponds to a different SR-wrapped dominance relation $\relation$ (crafted by modifying $\compRelationZ$), as defined in each row of Table~\ref{tab:allrel}.
For instance, for the \emph{allBest} option of the constrained strategy ($\compRelationStrict$), GOFOR relies on the relation $\compRelation$, that itself combines the relation $\compRelationZ$ and the simple scalar comparison $\leq$, to rely on the number of segments and so discriminate equal distances with respect to $\compRelationZ$.

From $\problemProperties$, constraints $\Constraints$, the SR-wrapping strategy and the diversity option leading to $\relation$, we can now formally define the SR-wrapped problem for a given source. 

\begin{Definition}
A problem $\problem(\problemProperties, \Constraints, \relation)$ consists in finding, $\forall v \in V$, the minimal lists of segments towards $v$, under the constraints $\Constraints$, verifying some properties $\problemProperties$, and that are non-dominated with respect to a given dominance relation $\relation$ (the SR wrapping of $\compRelationZ$).
\end{Definition}

GOFOR solves $\problem$, the SR-wrapped problem. 
More precisely, GOFOR is able to efficiently extend an algorithm designed for a given initial problem to solve $\problem$ (with $\relation$ and other parameters set to the operator needs among all options). 
The main challenges lie in the SR encoding and the extension of an extra subset of dominated distances, as introduced in Section~\ref{sec:illus}.
Our framework not only provides correct and optimal solutions, it retrieves them with an efficient path exploration avoiding superfluous computations.

\section{Two Main Ingredients: Encoding, Isotonocity}

\subsection{Encoding Distances \& Properties: Paths and Segment Lists}
\label{sec:encoding}

To solve $\problem$, it is necessary to compute the segment list(s) encoding the paths being explored during the initial exploration to properly take SR into consideration. 
In particular, a loose encoding scheme\footnote{We focus on loose encoding as we are interested in distances and properties rather than structure. Strict encoding schemes are straight-forward to design, do not offer minimal segment lists and can be found in the literature~\cite{AubryPhD}.} is necessary to translate the distances explored in order to guide the search accordingly and consider MSD. 
Unlike existing loose encoding schemes, our algorithm is not only efficient (as it does not directly rely on an SR-graph), but also handles multiple topologies (node segments of distinct types in practice).

\myparagraph{Greedily and Loosely Updating Segment Lists}
Our encoding algorithm (Algorithm~\ref{algo:encode} with its subroutine~\ref{algo:extend}) follows an efficient approach. 
Given a path $p$ as input (the one currently explored), it incrementally translates $p$ into a segment list $L$ using a greedy strategy. 
Initially, all possible segments able to encode the first edge of $p$ (which may be either $\Node$ or $\Adj$ segments) are stored in $\LastSeg$ (Line \ref{algo:encode:line: singleElt} in Alg.~\ref{algo:encode}). These segments are then extended to include more and more edges of $p$ as long as it is possible (\ie the segment lists verify the desired properties and match, or are better than, the path's metrics). The extension process is outlined in Alg.~\ref{algo:extend}. Segments that cannot be further extended correctly are removed from $\LastSeg$. If no more segments can be extended to include properly the new edge $e$, a new segment is required. Among the remaining segments failing at $e$, one of them, say $S$, is added to the segment list $L$~\footnote{All segments lists can still be retrieved during the backtracking post-processing phase used to re-construct the desired set of segment lists (according to a given path diversity option).}, and $\LastSeg$ is reset with all possible segments that can encode edge $e$ (Line \ref{algo:extend:line: singleElt} in Alg.~\ref{algo:extend}). While path $p$ is used to guide the search, the segments lists may include other paths as well, but only if they verify the same properties and are equal or better than $p$ with respect to its distances.
While our approach encodes logical characteristics rather than structural ones, $p$ is nevertheless part of the set of paths encoded by the resulting segment list. 

\begin{algorithm}
  \caption{ \footnotesize \textsc{Encode}($G, \problemProperties, p$)}\label{algo:encode}
  \footnotesize
  $L$ := $[\ ]$\\

  $e$ := first edge of $p$\\
  $\LastSeg$ := $\{S = \Seg(t, e) \;|\;$
  ${e\in S} \wedge {d(S) = d(e)} \wedge \problemProperties(G, S), 
  \forall t\in \NodeTypes \}$\label{algo:encode:line: singleElt}\\
  \For{$e\in p[1{:}\ ]$}{
    $L, \LastSeg$ := \textsc{Extend}$(L, \LastSeg, e)$\\
  }
  Let $S$ be any segment in $\LastSeg$\\
  \Return $L\oplus S$ 
  
 \end{algorithm}


\begin{algorithm}
  \caption{\footnotesize \textsc{Extend}$( L, \LastSeg, e)$}
  \footnotesize
  \label{algo:extend}

  $\NewLastSeg$ := $\emptyset$\\
  \For{$S \in \LastSeg$}{
    \If{${S \circ e\neq \emptyset} \wedge {d(S \circ e) = d(S) + d(e)} \wedge \mathit{Properties}(S\circ e)$\label{algo:extend:line: extend S}}{
      $\NewLastSeg := \NewLastSeg \cup \{S\circ e\}$\\
    }
  }

  \uIf{$\NewLastSeg = \emptyset$\label{algo:extend:line: if NewLastSeg = emptyset}}
  {
    Let $S$ be any segment in $\LastSeg$\\
    $L := L \oplus S$\\
    $\LastSeg$ := $\{S = \Seg(t, e) \;|\;$
    ${e\in S} \wedge {d(S) = d(e)} \wedge \problemProperties(G, S), 
    \forall t\in \NodeTypes \}$\label{algo:extend:line: singleElt}\\
  }
  \Else
  {
    $\LastSeg := \NewLastSeg$\\
  }
  \Return $L, \LastSeg$

\end{algorithm}

Algorithm~\ref{algo:encode} returns a minimal encoding (Theorem~\ref{thm:encode}). 
To ease the reading all proofs are given in the appendix, we here only provide lemmas to highlight their constructions.
In particular, Lemma~\ref{lem:single segment isotonic} states that loose encoding is \emph{isotonic}, \ie when a segment loosely encodes a path, then a restriction of this segment loosely encodes a sub-path. 

\begin{theoremEnd}[end, restate, no link to proof]{Lemma}
  \label{lem:encode function loosely encodes}
  \Encode($p$) returns a loose encoding of $p$.
\end{theoremEnd}
\begin{proofEnd}
  We prove by induction the following loop invariant: at the end of the $l$-th iteration, for any segment $S$ in $\LastSeg$, $L \oplus S$ is a loose encoding of $p[0:l]$. 
  
  Assume that at the beginning of the $l$-th iteration $L$ is a loose encoding of $p[0:l']$, with $l'\in [0,l]$ and $\LastSeg$ is a set of segments that loosely encode $p[l':l]$.

  In particular $d(L) = d(p[0:l'])$ and $d(S) = d(p[l':l])$.

  In the function $Extend$, there are two cases:
  \begin{enumerate}
    \item If $\NewLastSeg = \emptyset$ in Line~\ref{algo:extend:line: if NewLastSeg = emptyset}, then the returned variable $L$ is a loose encoding of $p[0:l]$ by assumption ($L$ concatenated with a segment in $\LastSeg$). Moreover, the returned value of $\LastSeg$ contains only segments that loosely encode $p[l:l+1]$.
    \item Otherwise, by condition Line~\ref{algo:extend:line: extend S}, each segment $S\circ e$ in $\NewLastSeg$ loosely encodes $p[l', l+1]$. Indeed, $(i)$ $S\circ e$ contains $p[l', l+1]$ because $S$ contains $p[l', l]$ by assumption and $S\circ e$ contains $e=p[l: l+1]$ (by definition); and $(ii)$ it verifies $d(S\circ e) = d(p[l', l+1])$ because  $d(S) = d(p[l', l])$ by assumption and from the condition we have 
    \[
      d(S\circ e) = d(p[l', l]) + d(p[l: l+1]) = d(p[l', l+1]).
    \]
  \end{enumerate}

\end{proofEnd}

\begin{theoremEnd}[end, restate, no link to proof]{Lemma}
  \label{lem:single segment isotonic}
  Let $S=\Seg(\Stype, p[0], p[l])$ be a single segment that loosely encodes a path $p$ of length $l$. Then, for any $i,j$ in $[0,l]$, $i<j$, $\Seg(\Stype, p[i], p[j])$ loosely encodes $p[i:j]$.
\end{theoremEnd}
\begin{proofEnd}
  Recall that we required that $p[i:j] \in \Seg(\Stype, p[i], p[j])$, so we have to show that $d(p[i:j]) = d(\Seg(\Stype, p[i], p[j]))$.

  Let $k$ be any metric. By definition of segment 
  \begin{align*}
  d_k(\Seg(\Stype, p[0], p[i])) &\geq d_k(p[0:i]),\\ d_k(\Seg(\Stype, p[i], p[j])) &\geq d_k(p[i:j]),\\ d_k(\Seg(\Stype, p[j], p[l])) &\geq d_k(p[j:l]).
  \end{align*}

  So we have 
  \begin{align*}
    d_k(\Seg(\Stype, p[0], p[i])) & \\
    + d_k(\Seg(\Stype, p[i], p[j])) & \\
    + d_k(\Seg(\Stype, p[j], p[l])) &\geq d_k(p)
\end{align*}

On the other side, 
the second property of segments implies that any path in $\Seg(\Stype, p[0], p[i])$ can be extended to a path in $\Seg(\Stype, p[0], p[j])$ (extended through the edges of $p[i:j]$) and to $\Seg(\Stype, p[0], p[l])$ (extended through the edges of $p[j:l]$).
Thus by definition of the distance of a segment, the distance of $\Seg(\Stype, p[0], p[l])$ is at least the sum of distances of the partial segments \ie we have

\begin{align*}
    d_k(S) \geq &
    d_k(\Seg(\Stype, p[0], p[i])) + \\
    &d_k(\Seg(\Stype, p[i], p[j])) +\\
    &d_k(\Seg(\Stype, p[j], p[l]))
\end{align*}
  On the other side, we have
  $d_k(p) = d_k(S)$ by definition of loose encoding. 
So we obtain that the inequalities are in fact equalities, and we have 
\[
  d(\Seg(\Stype, p[i], p[j])) = d(p[i:j]) 
\]
So $\Seg(\Stype, p[i], p[j])$ loosely encodes $p[i:j]$

\end{proofEnd}

\begin{theoremEnd}[end, restate, no link to proof]{Theorem}
  \label{thm:encode}
  \Encode($p$) returns a loose encoding of $p$ with a minimal number of segments.
\end{theoremEnd}

\begin{proofEnd}
From Lemma~\ref{lem:encode function loosely encodes}, we know that the returned segment list $L$ is a loose encoding of $p$. We now prove that $L$ has a minimal number of segments. Assume by contradiction that there exists a segment lists $L'$ that is a loose encoding of $p$ and that has a smaller number of segments than $L = \textsc{Encode}(p)$.
Since $L'$ has fewer segments, there must be a segment $S'\in L'$ that encodes a subpath $p[l'_1:l'_2]$ of $p$ such that a segment $S\in L$ encodes $p[l_1:l_2]$ with $l_1\geq l_1'$ and $l_2<l_2'$.

Let $S'=\Seg(\Stype, p[l'_1], p[l'_2])$.
By Lemma~\ref{lem:single segment isotonic}, we know that $\Seg(\Stype, p[l_1], p[l_2+1])$ loosely encodes $p[l_1:l_2+1]$ (and it also verifies $\problemProperties$ as any sub-path does by assumption), which contradicts the fact that no segment that encodes $p[l_1:l_2]$ can be extended by the edge $p[l_2: l_2+1]$ (Line \ref{algo:extend:line: extend S}).
\end{proofEnd}

Our innovative encoding scheme seamlessly integrates with most path computation algorithms, making them aware of the number of segments needed for the paths they explore. By invoking these algorithms incrementally at each edge relaxation in practice, this method provides a highly efficient encoding scheme (no significant overhead as node segment information is retrieved in constant time). However, as detailed in Sec.~\ref{sec:illus}, maintaining the number of segments required during exploration is not sufficient: certain dominated distances must also be extended.

\subsection{An Extended Relation to Regain the Isotonicity}
\label{sec:iso}

As already showcased in the introduction, extending only the optimal segment lists (according to the chosen $\relation$) is not enough to find a solution to the SR-wrapped problem $\problem$. Indeed, the dominance function $\relation$ is not isotonic anymore. Although optimal solutions are not $\relation$-dominated (by definition), a prefix of an optimal solution \emph{could} be $\relation$-dominated, due to the peculiar way the number of segments evolves after an extension (as emphasized with the example of Fig.~\ref{fig:test})~\footnote{Interestingly, algorithms that directly explore the SR Graph do not require such a scheme. The transformation of the graph makes the SR metric isotonic and predictable by converting it into a simple hop count metric.}. 

\myparagraph{Extend the Dominance Relation $\relation$ to $\relationStrong$}
Our goal here is to define a new dominance relation, called \emph{extended} dominance relation, in order to regain the isotonicity property and define which distances should be extended to prevent compromising either the optimality or the efficiency of our framework. One possible but costly solution would indeed be to treat all the lists as non-dominated. While returning optimal and correct solutions, this method would result in a huge overhead. Our extended dominance function is not only correct, but limits the induced overhead to $\times |V|$ at worst when updating distances (at each edge relaxation). Such an overhead is minimal as any weaker solution would miss valid solutions.

Our extension is formally defined in the following definition.
Note that, for a segment $S$, $u \in S$ abusively means that there exists a path in $S$ passing through $u$.

\begin{Definition}\label{def:strong comp}
  Let $\relation$ be a dominance relation over SR-distances. The extended relation $\relationStrong$ associated with $\relation$ is defined over segment lists as follows. For any segment lists $L$ and $L'$, we have $L \,\relationStrong\, L'$, \ie $L'$ is $\relationStrong$-dominated by $L$, if either
  \begin{itemize}
    \item[$(i)$] $d(L) \,\relation\, d(L') \quad\land\quad L_{last}^{type} = L_{last}'^{type} \quad\land\quad L_{last}^{src} \in L'_{last}$
    \item[$(ii)$] or $d(L) \,\relation\, (d_0(L')-1, d_{\setminus 0}(L'))$
  \end{itemize}
  where $L_{last}$ denotes the last segment of $L$ and $L_{last}^{src}$ its source. Notation $d_{\setminus 0}$ refers to the exclusion of metric $d_0$ in the vector of considered metrics.

\end{Definition}

This extended relation applies directly to segment lists and not to distances. 
It is used to check whether a given relation could hold after the segment lists are extended, and ensure that only segment lists that are not currently $\relation$-dominated or could potentially \emph{become} non-dominated are extended. 
Informally, $L'$ is $\relationStrong$-dominated by $L$ if it is dominated by $L$ regarding relation $\relation$ and either $(i)$ the source of the last segment of $L$ is contained within the paths encoded by $L'_{last}$ (also considering the same type of last segments), or $(ii)$, $L$ does have strictly fewer segments. Indeed, if segment list $L'$ verifies case $(i)$, one can show that if $L'_{last}$ can be extended by an edge $e$, so can $L$, meaning that $L$ will remain better than $L'$ (see Lemma~\ref{lem:if Ssrc' in S then S extende => S' extended} and its proof).

This isotonic property of this extended relation is stated in Theorem~\ref{thm:strong comp is isotonic} (for any relation $\relation$ wrapped around $\compRelationZ$ or $\compRelationStrictZ$), after exhibiting said Lemma. Again, the associated proofs are given in the appendix.

\begin{theoremEnd}[end, restate, no link to proof]{Lemma}\label{lem:if Ssrc' in S then S extende => S' extended}
Let $S$ and $S'$ be two segments such that $\Ssrc['] \in S$, $\Sdest[']=\Sdest$, and  $\Stype[']=\Stype$. We have that, if $S$ can be extended by an edge $e$, then $S'$ can also be extended by $e$.
\end{theoremEnd}
\begin{proofEnd}
  Consider a path $p\in S$ that passes through node $\Ssrc[']$.
  The Lemma follows from the two properties of segments. Indeed, if $S\circ e$ is not empty, then, by the second property, $p\circ e$ is in $S\circ e$. By the first property the sub path of $p\circ e$ starting from $\Ssrc[']$ is in $S'\circ e$ so $S'\circ e$ is not empty.

\end{proofEnd}

\begin{theoremEnd}[end, restate, no link to proof]{Theorem}\label{thm:strong comp is isotonic}
    The extended dominance $\extended{\relation}$ is isotonic, \ie the extension of an $\extended{\relation}$-dominated path remains $\relation$-dominated, even after extensions.
\end{theoremEnd}
\begin{proofEnd}

For simplicity, we assume that $\relation$ is the relation $\compRelationStrict$.

Let $L$ be a segment list that is $\extended\compRelationStrict$-dominated by a segment list $L'$. Let $S$, resp. $S'$ be the last segment of $L$, resp. $L'$. 
Let us denote $L^{+1}$, resp.  $L'^{+1}$, the segment list $L$, resp $L'$, once extended by an additional edge $(u,v)$.

By assumption, $L' \extended\compRelationStrict L$, which means, either \\$(i)$ $d(L') \compRelationStrict (d_0(L) - 1, d_{\setminus 0}(L))$ or \\$(ii)$ $d(L') \compRelationStrict d(L)$,  $\Stype = \Stype[']$, and $\Ssrc['] \in S$. 
In both cases, we have $L' \compRelationStrict L$, hence we have $d(L')\compRelationStrictZ d(L)$.

We know that, after the extension, the distance of the segment list $L$ increases by $(\delta_S,d_{\setminus 0}(u,v))$, where $\delta_S$ is 0 or 1 depending on whether $S$ can be extended by $(u,v)$ or not. So the $\compRelationStrict$-dominance between $L^{+1}$ and $L'^{+1}$ depends only on $\delta_S$ and $\delta_{S'}$.

\textbf{Case \emph{(i)} and $S$ cannot be extended by $(u,v)$}:
Then, we have 
$\delta_{S'}\leq \delta_S = 1$ and $(i)$, which implies that\\$d(L'^{+1}) \compRelationStrict (d_0(L^{+1}) - 1, d_{\setminus 0}(L^{+1}))$, and in turn $L'^{+1}\extended\compRelationStrict L^{+1}$.

\textbf{Case \emph{(ii)} and $S$ cannot be extended by $(u,v)$}:
Then, either $\delta_{S'} = \delta_S - 1$ ($S'$ can be extended) and $(ii)$, which implies that $d(L'^{+1}) \compRelationStrict (d_0(L^{+1}) - 1, d_{\setminus 0}(L^{+1}))$, and in turn $L'^{+1}\extended\compRelationStrict L^{+1}$.

or $\delta_{S'} = \delta_S = 1$ and $(ii)$, which implies $d(L'^{+1}) \compRelationStrict d(L^{+1})$. Also, the last segment of $L^{+1}$ and $L'^{+1}$ are equal, so $L'^{+1}\extended\compRelationStrict L^{+1}$.

\textbf{Case \emph{(i)} and $S$ can be extended by $(u,v)$}:
If $S'$ can also be extended, then again, $\delta_{S'}=\delta_S = 0$ and $(i)$, implies $L^{+1}\extended\compRelationStrict L'^{+1}$. Otherwise, if $S'$ cannot be extended, $(i)$ still implies that $d(L'^{+1}) \compRelationStrict d(L^{+1})$, and since $S'$ is not extended, the source of last segment of $L^{+1}$ is node $u$ and in included in the last segment of $L'^{+1}$, so $L'^{+1}\extended\compRelationStrict L^{+1}$.

\textbf{Case \emph{(ii)} and $S$ can be extended by $(u,v)$}:
Then, since $S'^{\mathit{src}} \in S$ and $S'^{\mathit{type}} = S^{\mathit{type}}$, $S'$ can also be extended (by Lemma~\ref{lem:if Ssrc' in S then S extende => S' extended}), so we still have $d(L'^{+1}) \compRelationStrict d(L^{+1})$. Also, we still have $S_{\mathit{ext}}'^{\mathit{src}} \in S_{\mathit{ext}}$ and $S_{\mathit{ext}}'^{\mathit{type}} = S_{\mathit{ext}}^{\mathit{type}}$, where $S_{\mathit{ext}}$ and $S_{\mathit{ext}}'$ are the extension of $S$ and $S'$, respectively, so $L'^{+1}\extended\compRelationStrict L^{+1}$.
\end{proofEnd}

Thus far, with a peculiar attention to the computing complexity, we have shown how paths and their characteristics should be loosely encoded into segment lists (Sec.~\ref{sec:encoding}), and which extra segments lists should be extended in order to guarantee optimality with respect to $\relation$ (Sec.~\ref{sec:iso}). From these two ingredients, we now describe how GOFOR turns an initial algorithm solving a non-SR problem into its extended version, solving its related SR-wrapped problem $\problem$.  

\section{GOFOR-SR, an Efficient Recipe to solve $\problem$}
\label{sec:recipe}

Given a shortest path algorithm $\algo(\compRelationZ)$, looking for optimal paths with respect to a comparison relation $\compRelationZ$ (modeling an arbitrary initial problem), we present here our recipe to create $$\algo(Loose|Strict, \extended\relation)$$ 
That is an algorithm solving $\problem$, the SR-wrapped version of the initial problem, with a flexible configuration: the operator can plug SR with the most suited encoding paradigm (loose or strict) and SR-wrapped relation $\relation$ according to its needs.

\subsection{Wrapping SR Around $\algo(\compRelationZ)$}


We assume that $\algo(\compRelationZ)$ solves an additive routing problem -- possibly already multi-dimensional, based on an isotonic relation $\compRelationZ$, and overall aiming to optimize given distances, verify $\problemProperties$ and respect some constraints $\Constraints_{\setminus 0}$. 
GOFOR enhances $\algo(\compRelationZ)$ to take SR into account and solve $\problem(\problemProperties, \Constraints, \relation)$.

The crucial operation in all shortest path algorithms is to decide whether a path or rather its distance should be stored or not in the \textit{Priority Queue} (denoted PQ) to be extended later on. 
For any relation $R$, we denote by $e \inPQ{R} \mathcal{E}$ the operation of storing the element $e$ in the set $\mathcal{E}$ of all current optimal elements (with respect to the relation $R$). In more formal words, we have $e \inPQ{R} \mathcal{E} \Leftrightarrow \nexists e' \in \mathcal{E}$ such that $e' R~e$.

Note that the operation carried by $\algo(\compRelationZ)$ when extending a path $p$, and its distance, can be expressed as 

$$
d(p) \overset{\compRelationZ}{\extendable} \mathcal{D}
$$

This means that non-dominated distances with respect to $\compRelationZ$ are stored in the priority-queue $\mathcal{D}$ to be extended further.  

GOFOR replaces this operation in $\algo(Loose|Strict, \extended\relation)$ to 

$$
\looseEncoding(p) \overset{\overset{\lozenge}{\relation}}{\extendable} \mathcal{D_L},
$$

The segment list (loosely or strictly) encoding $p$ (as defined in Sec.~\ref{sec:encoding}) is stored within the priority-queue $\mathcal{D_L}$ to be extended further (the PQ, $\mathcal{D_L}$, is now segment-list-based
), if it is non-dominated with respect to the chosen extended relation $\overset{\lozenge}{\relation}$ defined in Sections~\ref{sec:notations} and~\ref{sec:iso}.

\subsection{Complexity of the Priority Queue Updates}

Overall, our main computational challenge lies in the performance of the PQ with our extended dominance relation, \ie the $\inPQRExtended$ operation. Since we consider a Dijkstra-like algorithm, the complexity overhead indeed comes from the management of the underlying PQ. In the worst case, the PQ contains an entire distance Pareto-front to all the destinations, \ie $n.\Gamma.c_0$ entries, where $n = |V|$. Observe that the PQ does not contain each non-dominated segment lists but only non-dominated distances, whatever the path diversity option. Segment lists having the same distance are grouped into a single entry. Thus, the complexity of the \texttt{extract\_min} operation (\ie retrieving the segment list to extend from the queue) is $O\left(n.\Gamma .c_0\times\log(n.\Gamma .c_0)\right)$.

Once extracted, each distance (and their underlying segment lists) in the Pareto-front are extended to all adjacent neighbors. We extend $O(m.\Gamma.c_0)$ distances, where $m=|E|$. The complexity of this extension depends on the number of segments lists: as the behavior of the extension depends on the last segment of the segment list, each one has to be extended individually, even if they share the same distance. The complexity to perform all extensions is $O(m.\Gamma.c_0.r)$ where $r$ is the maximum number of segment lists for a given distance. The number $r$ is bounded by $n\times|\NodeTypes|$, but empirical evaluation suggests that in practice, $1 \leq r < 2$, resulting in a very limited overhead.

After extending a distance and its underlying segment lists to a neighboring node, we compare them to the existing distances towards this node, keeping only non-dominated distances. Comparing each newly extended distance to the existing ones has a complexity of $O(m.(\Gamma.c_0)^2)$. If the new distance already exists towards the considered node, we merge the associated segment lists (the newly extended ones and the existing ones) into a single distance entry.

In summary, the worst-case complexity of GOFOR is in
\[
O\left(n.\Gamma.c_0 \times \log (n.\Gamma.c_0) + m.\Gamma.c_0.(\Gamma.c_0 + r) \right)
\]

Note that with the \emph{lex} strategy, the term $c_0$ can be ignored in the complexity as such a strategy does not plug the SR metric as a constraint in the problem. Thus, if the initial problem is not multi-dimensionnal in itself ($\Gamma = 1$), one can just ignore the term $\Gamma.c_0$ overall.

\subsection{A Meta-DAG to Perform Source-Based Load Balancing}

The last remaining challenge is to effectively utilize the set of computed solutions thanks to a condensed data structure. Most shortest path algorithms return a shortest path DAG, from which all optimal solutions can be easily reconstructed and used by routers.

In our case, representing the set of solutions through a \emph{logical} (segment-lists level) DAG, with edges representing segments instead of physical edges, is also suitable for structuring the computed optimal segment lists. However, subtleties arise when considering multi-metric problems, where $\Gamma > 1$. In such cases, segment lists that are solutions for a node $v$, even if they have the same distance, can pass through an intermediary node $u$ with different intermediary non-dominated distances. This introduces ambiguity when reconstructing segment lists. To address this, extra information must be stored within the segment-list DAG to ensure proper and coherent reconstruction of the segment lists.

\myparagraph{Backtracking on a Logical DAG of Segment Lists}

To effectively present the calculated solutions towards a node $v$, we introduce a "meta-DAG" (named as such because each edge represents a segment list that may encode multiple paths, and each node now possibly supports multiple distances). In this representation, each node $u$ is replaced by at most $\Gamma$ nodes $u_0, \ldots, u_{\Gamma-1}$, with each node corresponding to an occurrence of $u$ in different solutions. In the worst case, $u$ appears in every non-dominated solution. A directed link between $u_{d}$ and $u'_{d'}$ represents a segment that is part of a solution, where $d$ and $d'$ are the intermediary distances of the solution at nodes $u$ and $u'$, respectively.

\begin{figure*}
\centering
   
\begin{tikzpicture}[yscale=1.7, xscale=1.7, -,>=stealth',shorten >=1pt,auto,node distance=3cm,
    thick,main node/.style={circle,fill=white,draw,font=\sffamily,inner sep=2pt},main node2/.style={circle,fill=white,draw,font=\sffamily,inner sep=5pt},
    subnode/.style={rectangle,font=\sffamily,text=white,fill=black,draw,inner sep=1.5pt},
    adj/.style={densely dotted},
    edgelabel/.style={fill=white,inner sep=1pt, anchor=center, pos=0.5,font=\sffamily}]

\node[main node, fill=black!20] (S) at (-0.7,0.7) {S};
\node[main node] (1)                at (1,1) {1};
\node[main node] (2)                at (1.8,1) {2};

\node[main node] (3)                at (-0.3,0) {3};

\node[main node] (4)                at (1.1,0) {4};
\node[main node] (5)                at (2,0) {5};

\node[main node,fill=black!20] (D)  at (2.6,0.5) {D};

\scriptsize
\node[main node] (6)                at (0.3,0.6) {6};
\node[main node] (7)                at (0.4,0) {7};

\draw (S) [] edge[] node[edgelabel] {2,1} (3);
\draw (3) [] edge[] node[edgelabel] {1,1} (6);
\draw (3) [] edge[] node[edgelabel] {4,1} (7);
\draw (3) [bend left=50] edge[] node[edgelabel] {4,1} (1);
\draw (6) [] edge[] node[edgelabel] {3,1} (7);
\draw (6) [] edge[] node[edgelabel] {2,1} (1);
\draw (7) [] edge[] node[edgelabel] {1,1} (4);

\draw (1) [] edge[] node[edgelabel] {2,1} (4);
\draw (1) [] edge[] node[edgelabel] {4,2} (2);
\draw (1) [bend left=20] edge[] node[edgelabel,pos=0.6] {1,5} (5);
\draw (4) [bend left=20] edge[] node[edgelabel] {2,2} (5);
\draw (4) [bend right=20] edge[] node[edgelabel] {3,1} (5);
\draw (4) [bend left=10] edge[] node[edgelabel,pos=0.4] {2,1} (2);
\draw (2) [bend left=10] edge[] node[edgelabel] {1,1} (5);

\draw (2) [] edge[] node[edgelabel] {3,1} (D);
\draw (5) [] edge[] node[edgelabel] {2,1} (D);

\normalsize
\begin{scope}[yshift=0cm,xshift=-4cm,scale=0.9]
\node[main node2, fill=black!20] (S) at (-0.55,0.48) {S};
\node[subnode] (S1) at (-0.3,0.46){}; 
\node[main node2] (1)                at (0.8,1.3) {1};
\node[subnode] (11) at (0.8,1.56) {\scriptsize 5,3};
\node[subnode] (12) at (0.8,1.08) {\scriptsize 6,2};

\node[main node2] (2)                at (1.8,1) {2};
\node[subnode] (21) at (1.7,0.76) {\scriptsize 9,5};

\node[main node2] (3)                at (0.4,0.2) {3};
\node[subnode] (31) at (0.4,0.46) {\scriptsize 2,1};

\node[main node2] (4)                at (1.1,0) {4};
\node[subnode] (41) at (0.85,-0.1) {\scriptsize 7,4};
\node[subnode] (42) at (1.3,0.2) {\scriptsize 8,3};
\node[main node2] (5)                at (2,0) {5};
\node[subnode] (51) at (2,-0.25) {\scriptsize 10,5};

\node[main node2,fill=black!20] (D)  at (2.55,0.77) {D};
\node[subnode] (D1) at (2.55,0.45) {\scriptsize 12,6};

\node[subnode] (00) at (3,1.5) {\scriptsize x,y};
\node at (3,1.25) {\scriptsize \emph{IGP,Delay}};

\scriptsize
\draw (S1) [bend left=20,in=140] edge[->] node[edgelabel] {5,3} (11);
\draw (11) [bend left=0,out=40, in=190] edge[adj,->] node[edgelabel] {4,2} (21);
\draw (21) [bend left=0] edge[adj,->] node[edgelabel,pos=0.4] {3,1} (D1);

\draw (S1) [] edge[->] node[edgelabel,pos=0.4] {2,1} (31);
\draw (31) [bend left=10] edge[adj,->] node[edgelabel,pos=0.4] {4,1} (12);
\draw (12) [bend left=10] edge[->] node[edgelabel,pos=0.23] {2,1} (42);
\draw (42) [bend left=0] edge[->] node[edgelabel] {4,3} (D1);

\draw (S1) [bend right=30,out=-60] edge[->] node[edgelabel] {7,4} (41);
\draw (41) [bend left=10, out=60,in=140] edge[->] node[edgelabel,pos=0.26] {2,1} (21);
\draw (41) [bend right=10,out=-60] edge[adj,->] node[edgelabel] {3,1} (51);
\draw (51) [bend right=30, out=-60] edge[->] node[edgelabel] {2,1} (D1);

\end{scope}

\begin{scope}[yshift=1.4cm, xshift=-2.5cm]
\scriptsize
\draw (0,0.2) [] edge[->] node[edgelabel,pos=0.4] {Node} (0.7,0.2);
\draw (0,0) [] edge[adj,->] node[edgelabel,pos=0.4] {Adj} (0.7,0);
\end{scope}


\end{tikzpicture}
\caption{A raw network (at the right), and its resulting meta-DAG of \emph{all} the segment lists from S towards node D having distance $(12,6)$.
}
\label{fig:metadag}
\end{figure*}
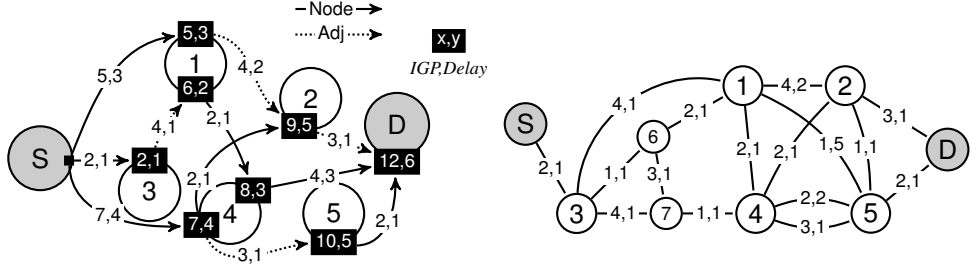

The meta-DAG serves as a tool to retrieve one or all segment lists from the set of solutions. In Figure \ref{fig:metadag}, we provide an illustration of such a meta-DAG along with its initial network for a complex use-case, DCLC-SR (wrapping SR on DCLC). The objective is to retrieve all the best IGP paths, encoded with their best segment lists, from node \nodetikz{S} to node \nodetikz{D} while adhering to a delay constraint of 7 ms, for example. The DCLC problem itself generates multiple non-dominated optimal solutions with its two initial metrics.

In the meta-DAG on the left, we represent the node and adjacency segments used between nodes. Some nodes can be reached with distinct non-dominated distances (because DCLC is itself a multi-metric problem): for instance the physical node \nodetikz{4} has two corresponding nodes in the meta-DAG: $4_{(7,4)}$ and $4_{(8,3)}$. In fact, there exists several optimal segment lists passing through \nodetikz{4} to reach \nodetikz{D}, \eg \linktikzzzzz{S}{3}{1}{4}{D} and \linktikzzzz{S}{4}{5}{D}. This structure, thanks to the added information described, allow reconstructing the desired segment lists.

In practice, this meta-DAG can be built by backward induction from the destination \nodetikz{D} to the source \nodetikz{S}, by listing, for all segments lists ending at $u$ and with distance $d$, the source of their last segment (which become the predecessors of $u_d$ in the meta-DAG).

In practice, this Meta-DAG could be used by the edge router, the source itself, to perform source-routed load-balancing across segment lists, \eg by choosing among optimal solutions through random walks in the Meta-DAG for each given flow. Such a feature allows to load-balance the traffic even if the latter is subject to complex traffic-engineering requirements (taking into account each node having multiple successors on the lists).
Further studies on the possibilities offered by such advanced load-balancing methods, and practical investigations regarding its feasibility directly in the data-plane of real hardware/software, is left for future work.

\newcommand{\delayConstrain}{\ensuremath{c_{\mathit{del}}}\xspace}

\section{SR-Wrapped TI-LFA, Least-Delay \& DCLC}
\label{sec:usecase}

In this section, we use GOFOR to wrap SR around three problems and algorithms: TI-LFA (a Fast-ReRoute, FRR, use-case), Least-Delay (LD) and Delay-Constrained-Least-Cost (DCLC). 
We aim to demonstrate the genericity and performance of our framework.
In particular, we focus on the evaluation of the efficiency of GOFOR on DCLC (the most computationally complex use-case). The source code of GOFOR and the experimental materials are available online~\cite{sourcecode}.

\subsection{Performance Analysis: Three Use-cases, a Single Recipe}

\myparagraph{Examples of General Settings for Defining $\problem$}
We consider three use-cases (DCLC, FRR, and LD) with the objective of transforming these problems into their corresponding SR-wrapped versions (DCLC-SR, FRR-SR, and LD-SR, respectively). As an example of strategy, let us first assume that the operator aims to ensure the deployability of the retrieved segments, indicating a preference for the \emph{Constrained} alternative to deploy SR. Additionally, the operator expresses an interest in obtaining all optimal solutions, implying that the \emph{allBest} diversity option should be used. Therefore, the operator should set the relation $\relation$ to $\compRelationStrict$ (refer to Table \ref{tab:allrel}), and implement the associated extended relation $\overset{\lozenge}{\relation}$ to compare the segments lists.

For the \textbf{DCLC-SR use-case}, where the initial goal is to find paths towards any destination with a delay bounded by \delayConstrain and minimizing the IGP distance, the related SR-wrapped problem can be formulated as: optimally encoding the DCLC paths requiring fewer than $MSD$ segments. One would set $\Constraints=(MSD,\infty,\delayConstrain)$ and leave $\problemProperties$ empty. With such settings, the desired $\compRelationStrict$ is expressed as follows:

\begin{align*}
    &(nbSeg_1, cost_1, delay_1) \compRelationStrict (nbSeg_2, cost_2, delay_2)\\
\Leftrightarrow & \left\{
\begin{array}{l}
    nbSeg_1 \leq nbSeg_2 \wedge cost_1 \leq cost_2 \wedge delay_1 \leq delay_2\\
    nbSeg_1 < nbSeg_2 \vee cost_1 < cost_2 \vee delay_1 < delay_2
\end{array}\right.
\end{align*}

For the \textbf{FRR-SR use-case}, the delay metric should be ignored\footnote{Mimicking a standard FRR use-case. GOFOR can also return DCLC-SR or LD-SR solutions avoiding a failed link.}, and the IGP cost should be optimized. However, the predicate $Properties(S)$ should be considered false if a path in $S$ uses the given failed link. The constraint should be set to $\Constraints=(MSD,\infty)$.

For the \textbf{LD-SR use-case}, the IGP cost should be ignored, and no specific $\problemProperties$ are required, as the only objective is minimizing the delay. Constraints should be set to $\Constraints=(MSD,\infty,\infty)$.

In practice, the modifications required for path computation algorithms are relatively light, involving only the implementation of the encoding scheme and the choice of the path comparison function to the relation $\overset{\lozenge}{\relation}$, as associated with the chosen $\relation$ (and adjusted with the wrapping strategy and its options).
While this section provides only a sample of possible use-cases, its purpose is to illustrate how GOFOR can transform almost any path computation problem into its SR-wrapped version, accommodating the operator's requirements in terms of optimization strategies and path diversity options.
 
We will now proceed to evaluate GOFOR on our three use-cases, considering several strategies and path diversity options.

\myparagraph{A SAMCRA-Based Implementation for GOFOR}

Recall that GOFOR, is a framework that transforms an existing algorithm to handle SR. Since we tackle use-cases encompassing several metrics, we used a generic multi-metric shortest path algorithm as a basis. 
We decide to rely on SAMCRA~\cite{VANMIEGHEM2001407} for its flexible PQ implementation. Although SAMCRA can return DCLC paths, it does not support any encoding paradigm nor extended comparison relation. We thus modified it according to our framework. The resulting code for all use-cases is available online~\cite{sourcecode}.

\myparagraph{Lattices and Realistic Topologies}

In the following, we start with an analysis highlighting the advantages of the loose encoding paradigm. We use specific graphs having two key properties: strong resilience and coarse valuation metrics that promote ECMP. These graphs are modified lattices with redundant links, created by doubling each link in a King's graph with a $0.3$ probability. The metrics assigned to the graph weights are uniformly random, with five possible values ranging from 1 to 5. 
These topologies are not intended to replicate real-world scenarios but rather to emphasize the benefits (reduction in the number of segments) of the loose encoding in extreme cases. 
The second evaluated criteria is the computation time performance (on DCLC-SR). 
For this, we use both the synthetic lattices described, and realistic topologies.
The latter are primarily extracted from the REPETITA framework~\cite{gay2017repetita}.

Experiments were conducted on a MacBook Pro Laptop, equipped with an Apple M1 Pro Chip and 16GB of RAM.

\subsection{Encoding Paradigms: a Look at the Number of Segments}

\begin{figure}
    \centering
    \includegraphics[scale=0.5]{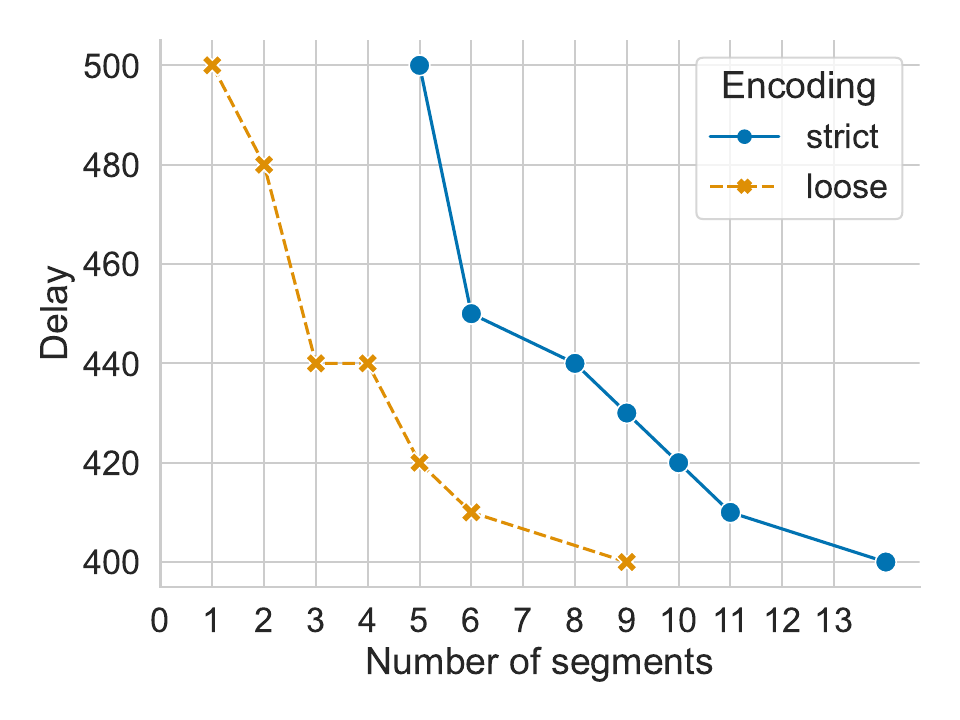}
    \caption{Comparison of the Pareto front solutions when performing loose encoding and strict encoding for LD-SR (with Cons.allBest).}
    \label{fig:segmentListLength1}
\end{figure}

\begin{figure}   
    \centering    
    \includegraphics[scale=0.5,trim={0 0 0 0.9cm},clip]{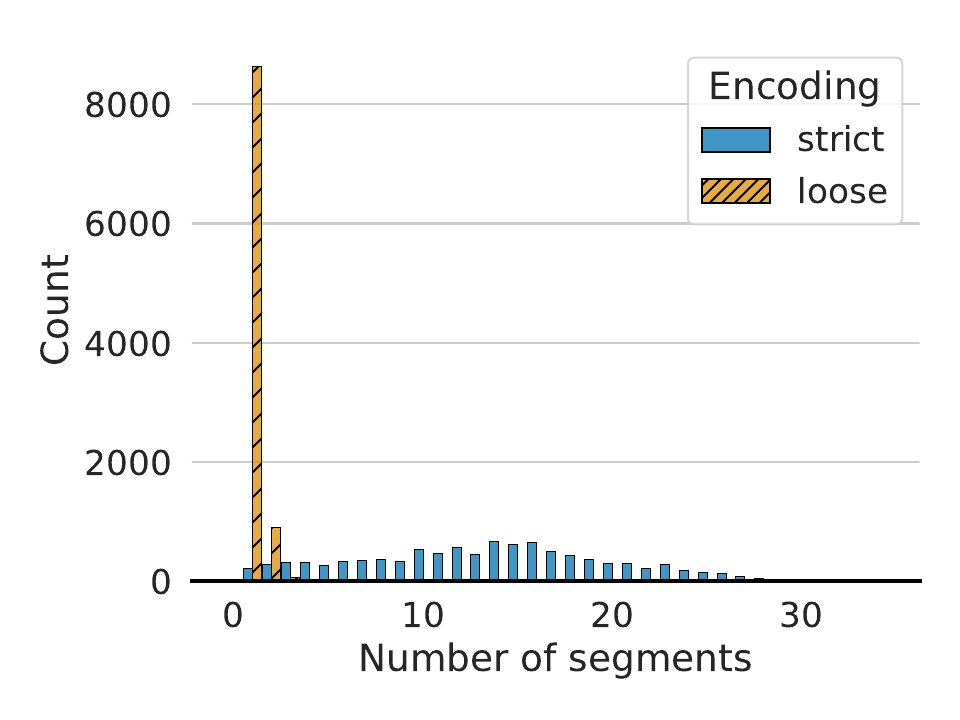}
    \caption{Distribution of the number of segments for the solutions found using loose encoding and strict encoding for FRR-SR (with Lex.all).}
    \label{fig:segmentListLength2}
\end{figure}

We first study how well our loose encoding leverages ECMP compared to strict encoding.
\myparagraph{Strict vs. Loose Encoding}
We compared the length of segment lists of the two paradigms in two use-cases. Figures \ref{fig:segmentListLength1} and \ref{fig:segmentListLength2} show the differences from these two perspectives. The first perspective illustrates the resulting Pareto Fronts for a specific source-destination pair in our graph set, focusing on LD-SR (with $\compRelationStrict$). Relying on loose encoding significantly reduces the overhead induced by pushing segments to the packet. 

The second figure conveys the same message but for FRR-SR, using the \emph{Lexicographical} strategy and \emph{all} diversity options ($\compRelationAllLex$). The figure shows the number of solutions (avoiding the failed link) for each number of segments, summed for all the destinations.

As our graphs have symmetric valuation, only between 2 or 3 segments are required \cite{ti-lfa} to avoid a link failure (we include the final destination segment, and consider that each basic local LFA requires an adjacency segment). This limit is indeed satisfied by our loose encoding scheme, while strictly encoded solutions require much more segments to encode unique backup paths (as ECMP is frequent). 
The loose encoding paradigm is thus necessary to retrieve \textit{all} feasible segments lists (\eg with MSD=10).


\subsection{Experimental Complexity: GOFOR-SR is Lightweight}

We now investigate how GOFOR performs compared to SR-graph-based frameworks with respect to the execution times of the transformed algorithm. We restrict here our study to DCLC-SR as it is the most challenging use case.

\begin{figure}
    \centering
    \includegraphics[scale=0.5]{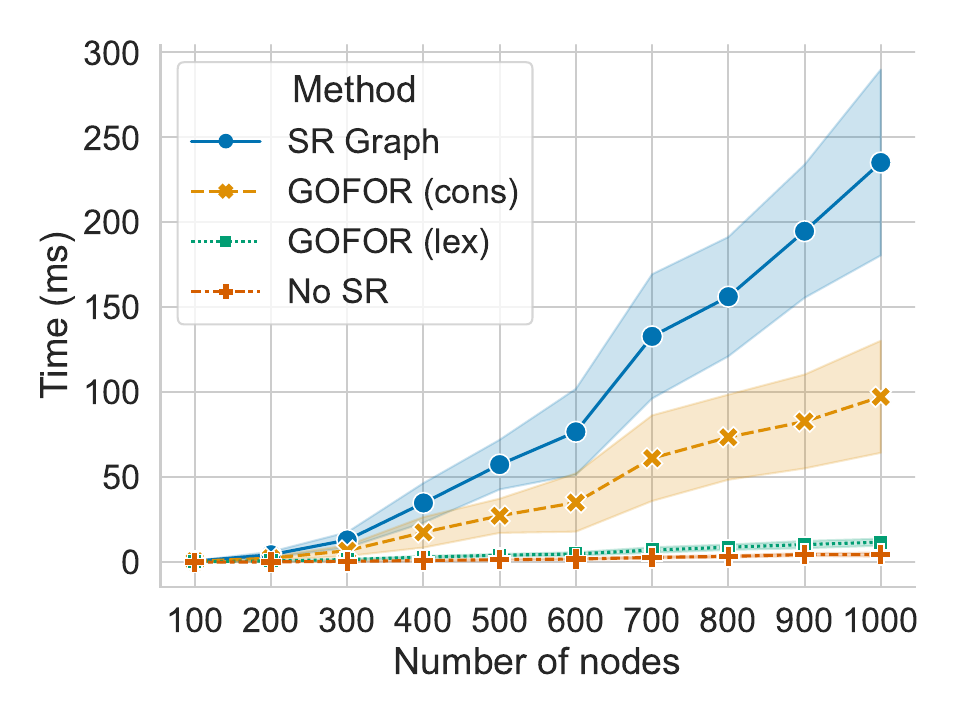}
    \caption{Comparison of the 
    execution time overhead to solve DCLC-SR.}
    \label{fig:Time}
\end{figure}

\myparagraph{A Limited Computing Time Overhead}
Figure \ref{fig:Time} provides a comparison of the computing time required by GOFOR with a \emph{Constrained} strategy ($\compRelationStrict$); GOFOR with a \emph{Lexicographical} strategy  ($\compRelationStrictLex$); SAMCRA as baseline (without SR); and SAMCRA on top of the fully meshed SR-Graph given as input. We rely on the \emph{allBest} diversity option in each of these setups.
We consider lattices of variable sizes to plot the computing time evolution according to their dimensions. The figure shows the average and standard deviation among 100 runs for each size.


\begin{figure}
    \centering
    \includegraphics[scale=0.5]{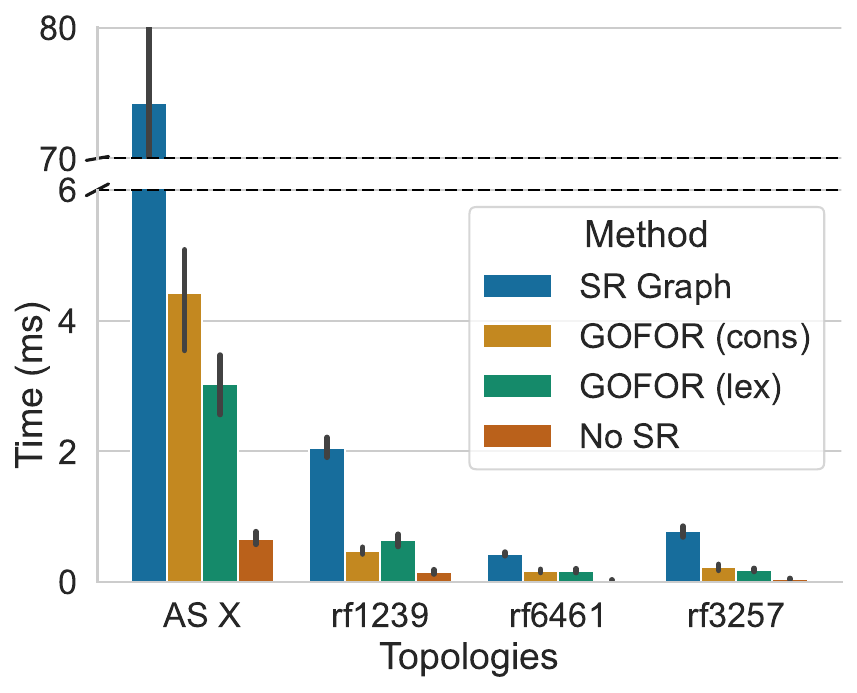}
    \caption{Execution times on realistic topologies of around \num{2000}, \num{1000}, \num{370} and \num{300} edges respectively.}
    \label{fig:realTime}
\end{figure}

Finally, we perform the same experience on realistic topologies from the REPETITA framework, on ASes ranging in size from 2000 edges to only 300. The results are shown in Fig~\ref{fig:realTime}. Note that there is a break in the y-axis of the figure.

Both figures show a clear tendency: GOFOR \emph{Lex.} has a negligible overhead (regarding the initial algorithm ignoring SR) and GOFOR \emph{Cons.} outperforms the best existing methods relying directly on the SR-Graph. While the performance of \emph{Cons.} compared to \emph{Lex.} looks significant at first glance, it is worth recalling that \emph{Cons.} fully adds the SR dimension to the initial problem (the MSD constraint is effective and cannot be ignored: $c_0 > 1$). 

Furthermore, this result is particularly obvious with the tested lattices, intentionally designed to amplify such effects (as $\Gamma$ becomes not negligible). However, on realistic topologies where problem instances are generally simpler, our SR-wrapped algorithm consistently solves this multi-metric problem in under 4ms. Experimental results on these realistic topologies not only showcase the superior performance of GOFOR compared to other frameworks but also demonstrate the minimal overhead introduced by our framework. We experimentally observe that $r < 2$ on average on our realistic cases, while we also only have $r < 3$ with our lattices.
We envision to theoretically investigate the distribution of $r$ in future works, \eg with an average complexity analysis on random graphs.

\section{Related Work}
\label{sec:relwork}

SR has generated a lot of traction, leading to contributions from both the industry and academia~\cite{SRSurvey}. It has been used to perform fine-grained monitoring~\cite{SCMon, 10.1145/3281411.3281426}, increase network resiliency~\cite{7145304, 10.1145/3281411.3281424} or perform traffic-engineering~\cite{WU2022}. Rather  than solving a specific routing use-case with SR, we proposed a general framework allowing to easily adapt existing (and future) path-computing algorithms to SR, streamlining the deployment of SR.

Some work aims to mitigate the limitations induced by MSD, making SR more scalable. For example, uSIDs carry several instructions in a single segment~\cite{uSID}. Similarly, binding segments (BSID) can be used to create a one-to-one mapping relationship between a segment and a segment list. These solutions can be used in conjunction with our schemes.  

Several SR-related contributions indirectly address the path encoding problem, although it is not their primary focus. Some use generic optimization frameworks to combine segments and create compliant segment lists~\cite{DEFO, 9796747}. Aubry proposed a dynamic programming approach, computing paths incrementally based on segment list size~\cite{AubryPhD}. Another option consists in exposing segments into a SR-graph, treating segments as edges, and execute algorithms on this inflated fully-meshed graph while limiting the exploration dept to MSD~\cite{LUTTRINGER2022109015, 7562080}, or just encoding a specific input path~\cite{Lazzeri2015EfficientLE}. While these techniques also support loose encoding, they suffer from performance drawbacks due to the density of the graph they explore. Exploring the SR-graph or using dynamic programming leads to $n^2$ operations while GOFOR mitigates this issue when the network graph is sparse, as it is often the case in practice. 
Other generic optimization frameworks generally need to heavily restrict MSD (often to 2 or 3) to reduce the exploration space.

Several works propose to strictly encode a specific path given as input to iteratively find the longest subpath encodable in a single segment~\cite{AubryPhD,7778603, 7313628,7417097}. 
While such schemes follow the same greedy approach as ours, these schemes are meant to be used \emph{a posteriori}, and thus neither leverage ECMP, nor ensure that the segment list found is deployable and/or minimal with respect to the operator needs. Finally, although the basic principles of GOFOR were mentioned in our previous work~\cite{LUTTRINGER2022109015}, the latter were neither complete nor were they proven, and were not generic nor evaluated. In this paper, we also add the support of distinct strategies, multi-topology and several path diversity models (enabling so fine-grained source-controlled load balancing) for various use-cases.

With GOFOR, we tackle the path encoding problem in a generic \emph{and} optimal fashion, considering nearly all basic use-cases and offering many key operational features. By tying together the paths and the segment lists computation, GOFOR returns all the relevant solutions. GOFOR performs better than concurrent approaches as it leverages the sparsity, and overall characteristics, of real IP networks as shown in Sec.~\ref{sec:usecase}.

\section{Conclusion}
The conventional best-effort routing paradigm, while scalable, falls short of meeting all the requirements of IP networks. Certain use-cases demand deviations from basic shortest routes to navigate failures or consider multiple metrics as additional constraints and objectives. Segment Routing (SR) stands out as one of the most popular options for deploying flexible routes loosely guided from the source.

However, the introduction of SR adds an operational metric, the number of segments (or detours), and a new challenge, efficiently retrieving all optimal segment lists. Minimizing these detours is crucial to ensure line-rate speed packet processing. Existing encoding schemes, which convert paths to segment lists, often overlook this aspect. They either convert paths a posteriori and/or fail to leverage Equal-Cost Multi-Path (ECMP), resulting in inflated segment lists. Alternatively, they impose a quadratic overhead in the number of nodes by relying on a complete SR-graph.

In this paper, we introduce a novel approach that enhances existing algorithms to return optimal and deployable segment lists instead of paths. Our initial step involves the design of a \emph{loose encoding} scheme, which seamlessly integrates into current path computation algorithms. This scheme computes minimal segment lists for the paths being explored, utilizing all available segment types and leveraging ECMP. We then address the transformation of the path comparison relation to ensure the discovery of the optimal segment list, despite the loss of isotonicity (\ie substructure optimality) induced by the SR metric. The newly proposed relation, able to accommodate various optimization strategies and path diversity options based on the operator's requirements, minimizes the computation overhead to a strict minimum. We show that algorithms extended by our resulting framework, GOFOR, remain particularly efficient compared to existing SR-aware approaches, even facing challenging multi-criteria routing problems.


As a promising future work, we aim to investigate whether our meta-DAG of optimal solutions could be fitted into the data-plane directly to perform efficient source-driven, TE-aware load-balancing across all segment lists.

\bibliographystyle{IEEEtran}
\bibliography{ref.bib}

\appendix
\section{Omitted Proofs}
\printProofs
\end{document}